\documentclass[aps,prb,amsmath,amssymb,preprintz,twocolumn]{revtex4-2}

\usepackage{graphicx}
\usepackage{amsmath}
\usepackage{braket}
\usepackage{amssymb}
\usepackage{bbm}
\usepackage{bm}
\usepackage{mathtools}
\usepackage{hyperref}
\usepackage{natbib}
\usepackage{mathtools}\usepackage[normalem]{ulem}
\graphicspath{{Figures}}
\usepackage[normalem]{ulem}

\usepackage[x11names]{xcolor}

\begin{document}
	
	\title{Lifshitz transitions and   Weyl semimetals \\from a topological superconductor  with    supercurrent flow}
	\author{Fabian G. Medina Cuy}
		\author{Francesco Buccheri}
  \author{Fabrizio Dolcini}
	\affiliation{Department of Applied Sciences and Technology, Politecnico di Torino, corso Duca degli Abruzzi 24, 10129 Torino (Italy)}
	\date{\today}
	
	\begin{abstract}A current flowing through a superconductor induces  a spatial modulation in its superconducting order parameter, characterized by a wavevector $Q$  related to the total momentum of a Cooper pair. Here we investigate this phenomenon in a $p$-wave topological superconductor, described by a one-dimensional   Kitaev model. We demonstrate that, by treating $Q$ as  an extra synthetic dimension,  the current carrying non-equilibrium steady state can be mapped into the ground state of a half-filled two-dimensional  Weyl semimetal, whose Fermi surface  exhibits Lifshitz transitions when varying the  model parameters. Specifically, the transition from   Type-I to Type-II Weyl phases corresponds to the emergence of  a gapless $p$-wave superconductor, where Cooper pairs coexist with unpaired electrons and holes. 
Such transition is signaled by the appearance of a sharp cusp in the $Q$-dependence of the supercurrent,    at a critical value $Q^*$ that   is robust to variations of the chemical potential $\mu$. We determine the maximal current that the system can sustain in the topological phase, and discuss possible implementations.  
	\end{abstract}

	\maketitle

\section{Introduction.}
Weyl semimetals (WSMs) and Topological superconductors (TSs)  might lead to an actual breakthrough in quantum science and technology.

Indeed WSMs are quite   promising   for applications in ultra-fast electronics  and photonics  due to their  peculiar linear band spectrum and large carrier mobility \cite{Shekhar2015,Armitage2018}. These materials    
are commonly divided into two families, dubbed Type-I \cite{Burkov2011,Hasan2017} and Type-II \cite{Soluyanov2015,Wang2016,Li2017,yao2019}, depending on the    tilting of the cone characterizing their   electronic spectrum near special point nodes. In turn, these bulk nodes also protect the existence of topological Fermi arc states on the WSM surfaces \cite{Bovenzi2018,Breitkreiz2019,Buccheri2022t}, as confirmed by various ARPES experiments \cite{Xu2015,Wu2016,Tamai2016,Morali2019}. While three-dimensional WSMs have been vastly studied, more recently a growing interest is devoted to  two-dimensional WSMs, \cite{Hao2016,Irobe2016,guo2019,He2020,Meng2021}, also in view of their possible realization with cold atoms\cite{guo2019}.

Similarly, TSs are materials with a huge potential in applications, as they combine two remarkable properties. On the one hand, they host edge  modes, known as Majorana quasi-particle (MQPs)
\cite{alicea2012,aguado2017,zhang2018,dassarma2023,aghaee2023}, featuring peculiar non-local correlations and unconventional braiding properties that could be harnessed for topologically protected  quantum computation. On the other hand, they exhibit a  dissipationless transport  that is ideal to develop  green nanoelectronics. 
 Various implementations of 1D TSs, based e.g. on proximized spin-orbit nanowires\cite{lutchyn2010,oreg2010}, quantum spin Hall edge systems contacted to   ferromagnets\cite{fu-kane2008,fu-kane2009} and ferromagnetic  atom  chains\cite{Choy2011,nadjperge2013,simon2013,vonoppen2013,franz2013,kotetes2014}, have been proposed and are supported by promising, although  not yet conclusive, experimental confirmations.\cite{kouwenhoven2012,furdyna2012,heiblum2012,kouwenhoven2018,yacoby2014,yazdani2014,tang2016, yu2021non}  

\begin{figure}[h]
		\includegraphics[scale = 0.3]{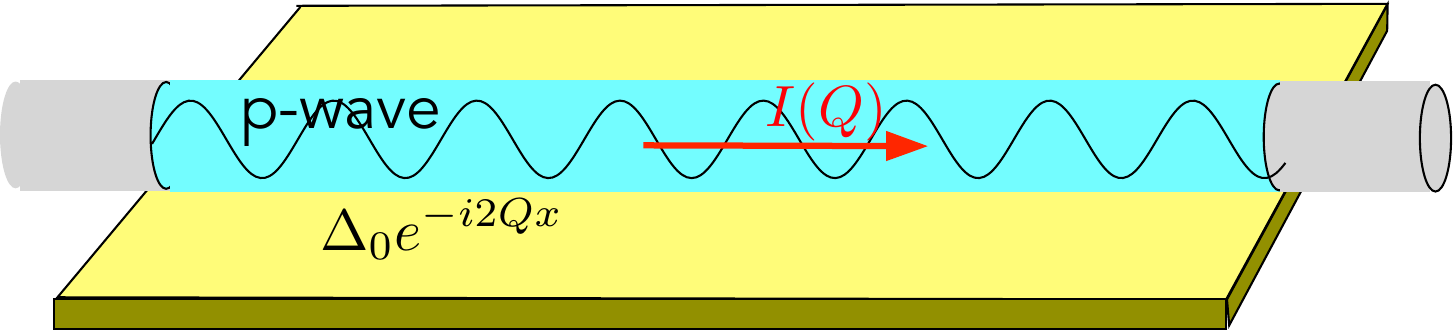}
		\caption{\label{Fig1}  A topological $p$-wave superconductor carrying a supercurrent flow. Its order parameter exhibits a phase modulation $Q$ related to the momentum of Cooper pairs.\\ }
\end{figure}

In this work, we show that the phase modulation $Q$ emerging in the   order parameter of a superconductor in the presence of a current flow\cite{tinkham-book} naturally provides an intriguing connection  between  TSs and WSMs.    
Indeed, by treating $Q$ as an additional synthetic dimension, we show that the nonequilibrium stationary state of the 1D superconductor with a current flow
can be mapped onto the ground state of a 2D half-filled fermionic model. As a consequence, the appearance of the various phases of the 1D superconductor can be understood as the result of a Lifshitz transition\cite{volovik2017,Volovik2018} in the Fermi surface of the associated 2D model.
In particular, a 1D $p$-wave TS is mapped onto a 2D WSM\cite{ohberg2011,zhang2015,guo2019,zhao2022,guo2023,zhang2023}.   By varying the model parameters, the  transition from type-I to Type-II WSM  corresponds  to the appearance of a gapless  superconducting phase in the TS, where Cooper pairs coexist with unpaired electrons and holes.
In turn, this also determines a  change in the range of $Q$-values where the gapped TS exists, which is controlled by its chemical potential or superconducting order parameter, depending on whether the associated WSM is in the Type-I or in the Type-II phase.
By exploiting the   mapping, we construct the low energy expression of a Fermi ``arc" in the 2D WSM and of the Majorana edge mode in the corresponding current-biased 1D superconductor.
By contrast,  a 1D $s$-wave superconductor corresponds to a 2D insulator or to a 2D conventional semimetal, depending on the parameter range. We show that such a difference between   $s$-wave and $p$-wave superconductors can be encoded in different topological classes  of closed circuits in the extended 2D Brillouin zone.

The effect of a superconducting phase modulation is typically neglected in most models  of Josephson junctions (JJs), with the argument that current conservation enables one to evaluate  the current in the normal weak link, whose properties ultimately determine the critical current.
This assumption has also been made in models for JJs\cite{nakhmedov2017,nakhmedov2020} or SQUIDs based on $p$-wave TSs.\cite{vonoppen2013,giuliano2016,lucignano2024}
However, because the phase modulation induced by the  current  modifies the bulk spectrum of the superconductor, the parameter range determining the  topological phases of a TS is actually affected by the supercurrent flow itself\cite{loss2019}. In particular, in the last two years the topological effects of a superconducting phase modulation has been investigated for the Kitaev model\cite{kitaev2001}, focusing on the regime where the magnitude $\Delta_0$ of superconducting order   parameter is larger than the bare bandwidth parameter $w$~\cite{yanase2022,ma-song2023}. Such a regime $\Delta_0>w$, however,   is hardly achievable in realistic implementations, and   an investigation in more realistic regimes is lacking. 
  
Our analysis overcomes this limitation. Indeed we investigate the effects of the superconducting phase modulation $Q$ on a $p$-wave  1D superconductor connected to reservoirs, as sketched in Fig.\ref{Fig1}. The system is   modelled by a  Kitaev   chain, characterized by a hopping strength $w$, a chemical potential~$\mu$, and   a magnitude $\Delta_0$ of the superconducting order parameter. By analyzing  arbitrary parameter values, we show that, while in the regime $\Delta_0>w$ the system remains always gapped  and the role of $Q$ is to merely modify the  values of $\mu$ separating the trivial from the topological phase, a richer scenario emerges for the physically relevant range $\Delta_0<w$. In particular, for $|\mu|<2\sqrt{w^2-\Delta_0^2} $ the phase modulation $Q$ can lead   the system to become a gapless $p$-wave superconductor. By computing the current in the Kitaev model as a function of the superconducting phase modulation $Q$, we show also that, because of the $p$-wave nature of the superconducting order parameter, the critical value   $Q^*$ determining the onset of the gapless phase is independent of the chemical potential  and only depends on the superconducting order parameter.
In our analysis, we determine an upper bound on current that can be driven through the system before the topological Majorana modes disappear.

The article is structured as follows. In Sec.\ref{sec2} we describe the model and its symmetries, whereas in Sec.\ref{sec3} we derive the general expression of the current carrying state  and determine in which range of parameter values it is a conventional gapped $p$-wave superconductor or a gapless $p$-wave superconductor.   Then, in Sec.\ref{sec4} we introduce the interpretation of $Q$ as a synthetic dimension and we show how the current carrying state of a 1D $p$-wave superconductor can be mapped onto the ground state of a 2D Weyl semimetal. Finally, in Sec.\ref{sec5} we derive the  current as a function of the phase modulation $Q$ and of the model parameters, while in Sec.\ref{sec6} we  summarize our results, and discuss possible implementations and future perspectives.

\section{Model and symmetries}
\label{sec2}
	{\it Model.} We consider a one-dimensional $p$-wave superconductor modelled as a Kitaev chain, whose Hamiltonian reads
	\begin{align}
		\begin{split}
			\mathcal{H}(Q) =& \sum_{j}\left\{ w\, \left(c^{\dagger}_{j}c^{}_{j + 1} +c^\dagger_{j+1} c^{}_j\right)-\mu\left(c^{\dagger}_{j}c_{j} - \frac{1}{2}\right) +\right. \\
			+& \left. \Delta_{0} \left(e^{-iQ\left(2j + 1\right)} c^{\dagger}_{j}c^{\dagger}_{j + 1} + e^{iQ\left(2j + 1\right)} c^{}_{j+1}c^{}_{j}\right) \right\}.
		\end{split}
		\label{Ham-real-space}
	\end{align}
	Here $c_{j} \, (c^{\dagger}_{j})$ corresponds to the annihilation (creation) operator at the lattice site $j$,  $\mu$ is the chemical potential, and $w > 0$ and $\Delta_{0} > 0$ are the magnitudes of the hopping amplitude and of the superconducting order parameter, respectively. The wavevector $Q$  (in units of the inverse lattice spacing) characterizing the spatial modulation of the order parameter, describes a net momentum $-2Q$ of the Cooper pair and  accounts for the presence of a current flowing through the system.

	
Since the phenomenon we aim to describe is a bulk effect, we can safely adopt the thermodynamic limit.  We assume that the number  $N_{s}$   of lattice sites is large, $N_{s} \gg 1$, and adopt periodic boundary conditions (PBCs) for the system. Thus, the superconducting spatial modulation in Eq.(\ref{Ham-real-space}), which is quantized as $Q = 2\pi n/N_{s}$ because of the PBCs,  can effectively   be treated as a continuum variable.  By introducing Fourier mode operators $c_k=N_{s}^{-1/2} \sum_j e^{-i k j} c_j$ and the Nambu spinors $\Psi_{k;Q} = (c^{\dagger}_{k-Q} ,  c^{}_{-k-Q} )^{T}$,   one can rewrite the Hamiltonian (\ref{Ham-real-space})  as 
	\begin{equation}
		\mathcal{H}(Q) = \frac{1}{2}\sum_{k \in {\rm BZ}}\Psi^{\dagger}_{k;Q} {H} (k;Q) \Psi_{k;Q},
		\label{Kitaev-v2}
	\end{equation}
where
	\begin{equation}
		{H}(k; Q) = h_{0} (k;Q) \sigma_{0} + \bm{h}(k; Q)\cdot \bm{\sigma},
		\label{Ham-BdG}
	\end{equation}
	is the Bogolubov de Gennes (BdG) Hamiltonian matrix, $\sigma_0$ is the $2 \times 2$ identity matrix, $\bm{\sigma} = \left(\sigma_{1},\sigma_{2},\sigma_{3}\right)$ are the Pauli matrices, and
  	\begin{align}
  h_{0}(k; Q)  =& 2w \sin{Q}\sin{k}\label{h0-def}\\
  \bm{h} (k; Q)  =&  \left(0,\, - \text{Im}\left\{\Delta (k)\right\} ,\, \xi(k;Q)\right) 
\label{h-vec-def}
	\end{align}
with 
   	\begin{align} 
		\xi (k; Q ) =& 2w \cos{Q} \, \cos {k}\,- \mu  \label{xi} \\
		\Delta\left(k \right) =& 2\Delta_{0}i \sin{k}. \label{delta}
	\end{align}

 {\it Symmetries.} By construction of the BdG formalism, the Hamiltonian (\ref{Ham-BdG})  fulfills the  particle-hole  constraint
	\begin{equation}
	\sigma_{1}H^{*}  (k; Q)\sigma_{1} = - {H}\left(-k;Q\right)\quad.
	\label{particle-hole}
	\end{equation}
The application of the  time-reversal transformation $\mathcal{T}\Psi_{k;Q}\mathcal{T}^\dagger=\Psi_{-k;-Q}$ (anti-unitary) and the spatial inversion transformation $\mathcal{I}\Psi_{k;Q}\mathcal{I}^\dagger=i\sigma_3 \Psi_{-k;-Q}$ (unitary) on the Hamiltonian Eq.(\ref{Kitaev-v2}), leads to the following relations for the BdG Hamiltonian (\ref{Ham-BdG})
	\begin{eqnarray}
		{H}^{*}(k;Q) = {H}\left(-k;-Q\right),
		\label{time-reversal} \\
  \sigma_3 H^{}(k;Q)   \sigma_3 = {H} (-k;-Q)\label{inversion} 
	\end{eqnarray}
showing that the presence of the spatial modulation $Q$ breaks both such symmetries.\\

 
\section{Gapped and gapless $p$-wave superconducting phases}
\label{sec3}
The normal modes of the quadratic Hamiltonian~(\ref{Kitaev-v2}),  and therefore its ground state and excitations, are determined by  diagonalizing the related BdG Hamiltonian~(\ref{Ham-BdG}). In order to describe the effects of the spatial modulation wavevector  $Q$ on the  superconducting state, two remarks are in order. 

First, we note that  $Q$ enters the BdG Hamiltonian~(\ref{Ham-BdG})  in a twofold manner. On the one hand, $Q$ appears in the third component  $\xi$   of   the $\bm{h}$-vector in Eq.(\ref{h-vec-def}). Such a term  represents a modification
\begin{equation}
\varepsilon(k) \rightarrow    \xi(k;Q)=\frac{\varepsilon(-k+Q)+\varepsilon(k+Q)}{2}
\end{equation}
of the bare dispersion relation $\varepsilon(k)=2w\cos{k}-\mu$ of the tight-binding model [first line of Eq.(\ref{Ham-real-space})], which results in the reduction $w \rightarrow w \cos{Q}$ of the  hopping parameter encoded in Eq.(\ref{xi}).
On the other hand, $Q$ introduces in the Hamiltonian~(\ref{Ham-BdG}) the additional $h_0$ term (\ref{h0-def}), which can be written as the difference
\begin{equation}
    h_0(k;Q)=\frac{\varepsilon(-k+Q)-\varepsilon(k+Q)}{2}
\end{equation}
between the bare energies of two electrons $(k,-k)$ in the Cooper pair frame $-Q$. Such a term is  odd in $Q$ and causes the breaking of time-reversal and spatial inversion symmetries of the model [see Eqs.(\ref{time-reversal}) and (\ref{inversion})]. \\
The second remark is that, since in Eq.(\ref{Ham-BdG}) the first term containing $h_0$ is proportional to the identity $\sigma_0$, the set of eigenstates of     Eq.(\ref{Ham-BdG}) is determined by the second term $\bm{h}\cdot \boldsymbol\sigma$ only.  However, because the term  $h_0  \sigma_0$ affects {\it the spectrum}, it also determines, for each~$k$, which single-particle eigenstate  is energetically more favorable and must be occupied. As a consequence, the actual many-particle state is modified by $Q$, and so are its topological properties. \\

\begin{figure*} 
		\includegraphics[scale = 0.58]{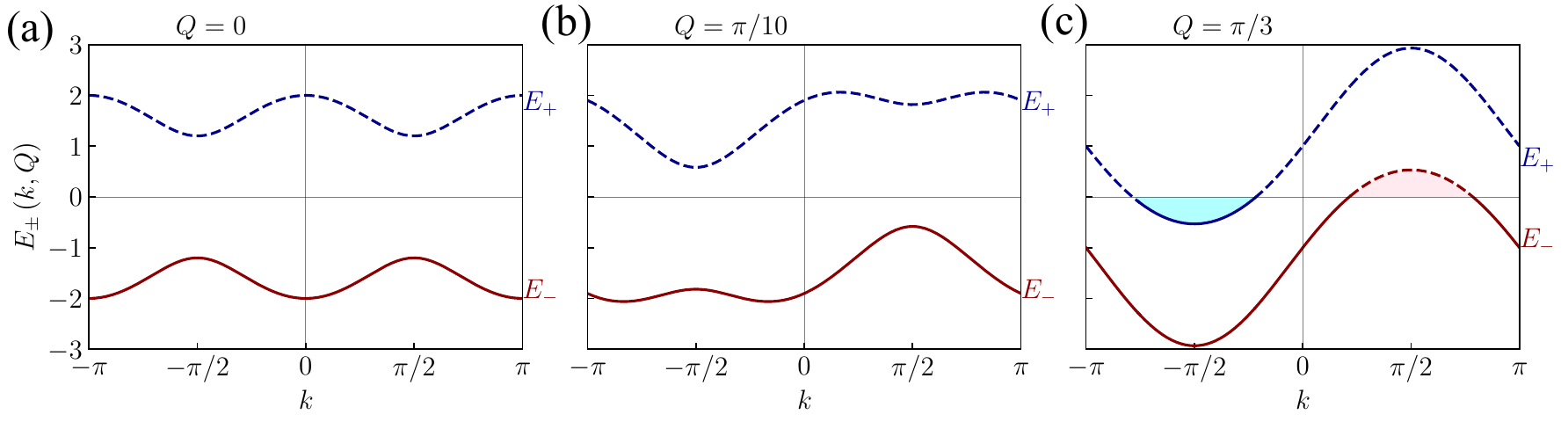}
		\caption{\label{Fig2} The spectrum of the upper band $E_{+}$ (blue dashed curve) and the lower band $E_{-}$ (red curve) for $\Delta_{0} = 0.6 w$, $\mu=0$. Panels (a), (b) and (c) refer to $Q=0$, $Q=0.1 \pi$ and  $Q=\pi/3$.}
	\end{figure*}

To describe in details  how this occurs, it is  worth  recalling  briefly the procedure determining the 
 the normal modes of the Hamiltonian~(\ref{Kitaev-v2}). 
	By means of the Bogolubov-Valatin unitary transformation
	\begin{equation}
		U_Q\left(k\right) = 
		\begin{pmatrix}
			u_Q\left(k\right) & - v^{*}_Q\left(k\right)\\
			v_Q\left(k\right) & u_Q\left(k\right)
		\end{pmatrix} 
	\end{equation}
 where 
	\begin{align}
		u_Q\left(k\right) =& \sqrt{\frac{1}{2}\left(1 + \frac{\xi(k;Q)}{h(k;Q)}\right)}\\
		v_Q\left(k\right) =& -i\,\text{sgn}\left(\sin\left(k\right)\right)\sqrt{\frac{1}{2}\left(1 - \frac{\xi(k;Q)}{h(k;Q)}\right)} \quad,
	\end{align}
 the BdG Hamiltonian (\ref{Ham-BdG}) can be brought to its  diagonal form $U^{\dagger}_Q {H} U_Q^{}  = {\rm diag}(E_{+},E_{-})$. The upper band $E_{+}$ and the lower band $E_{-}$ of the Kitaev model with superconducting modulation are given by   
 	\begin{equation}
		E_{\pm}\left(k;Q\right) = h_{0} (k;Q) \pm h(k;Q) \quad,
		\label{bands}
	\end{equation} 
 where   $h(k;Q)=|\bm{h}(k;Q)|=\sqrt{\xi^2(k;Q)+|\Delta(k)|^2}$. 
The above  Eq.(\ref{particle-hole}) implies that, for each $Q$-value, the two energy bands (\ref{bands}) are mutually related through the relation
 	\begin{equation}
		E_{-} (k;Q) =  - E_{+}(-k;Q) \quad,\label{rel_bands1} 
	 \end{equation}
 whereas  Eq.(\ref{time-reversal}) or Eq.(\ref{inversion})  imply that 
 	\begin{equation}
		E_{\pm}(k;Q) =    E_{\pm} (-k; -Q) \quad. \label{rel_bands2}
\end{equation}
showing that the presence of $Q$ makes the two bands no longer symmetric for $k \rightarrow -k$. 
 In Fig.\ref{Fig2}, the two bands are shown as a function of $k$, at fixed $\mu$ and $\Delta_0$ values, for three different values of $Q$. Panel (a) describes the customary $Q=0$ Kitaev model: The two bands are symmetric in $k$, the upper (lower) band is   positive (negative) $\forall k$, and a finite direct gap exists between the two bands. As one can see from panel (b) and (c), a finite $Q$ breaks the inversion symmetry [see Eq.(\ref{inversion})] and, when sufficiently large, it can also lead  the upper band $E_{+}$  to acquire  negative values, or equivalently, the lower band~$E_{-}$ to become positive for the opposite values $-k$. When this occurs, the gap closes {\it indirectly}.
 
While the physical consequences of this behavior will be discussed in the next subsection, here we note that the Hamiltonian (\ref{Kitaev-v2}) is straightforwardly brought into its normal modes as
\begin{equation}
		\mathcal{H}  = \sum_{k} E_{+} (k;Q) \left(\gamma^{\dagger}_{k-Q}\gamma^{}_{k-Q} - \frac{1}{2}\right)\quad,
		\label{diagonal}
	\end{equation}
where we have exploited Eq.(\ref{rel_bands1}) and we have introduced Bogolubov quasi-particles  $\Gamma_{k;Q}= (
			\gamma_{k -Q},
			\gamma^{\dagger}_{-k -Q})^T =
 {U}_Q^\dagger(k) \Psi_{k;Q}$, which fulfill   $\{ \gamma^{\left(\dagger\right)}_{k}, \gamma^{\left(\dagger\right)}_{k'}  \} = 0$ and $ \{ \gamma^{}_{k}, \gamma^{\dagger}_{k'}  \} = \delta_{k,k'}$ and are explicitly written as
	\begin{align}
		\begin{split}
		\gamma_{k-Q} =& u_Q\left(k\right)c_{k-Q} + v_Q^{*}\left(k\right)c^{\dagger}_{-k-Q} \\
		\gamma^{\dagger}_{-k-Q} =& - v_Q\left(k\right)c_{k-Q} + u_Q\left(k\right)c^{\dagger}_{-k-Q}
		\end{split}.
\label{Bogoliubov-quasi}
	\end{align}

As is well known,  in view of the redundancy of the Nambu spinor degrees of freedom, only one band is physically meaningful. This is why $\mathcal{H}$  is expressed in Eq.(\ref{diagonal}) in terms of $E_{+}$ only. However, since in the following we shall have to deal both with states at $E_{+}(k;Q)$ and $E_{+}(-k;Q)$, it will be useful to use the lower band $E_{-}(k;Q)$ as a notation for $-E_{+}(-k;Q)$, in view of Eq.(\ref{rel_bands1}).

	\subsection{The supercurrent carrying state}
	Equation~(\ref{diagonal}) expresses the Hamiltonian as a collection, labelled by $k$, of single fermionic states with energy $E_{+}$ that can be either occupied or empty. 
  For a given  value~$Q$ of the wavevector modulation, $E_{+}$ can in general take both positive and negative values as a function of $k$ [see Fig.\ref{Fig2}(c)]. Thus, one can partition the  BZ  in two regions, labelled $S_+$ and $S_{-}$ and defined as
 \begin{equation}
		\begin{array}{ccc}
			k \in S_{+}  & \mbox{iff} &E_{+}(k;Q) > 0   \\
			k \in S_{-} & \mbox{iff} &       E_{+}(k;Q) < 0
		\end{array}\quad.
	\label{first_partition}
	\end{equation}  
From Eq.(\ref{diagonal}) we deduce that,   if $k \in S_{+}$, it is energetically more convenient to leave the $k$-th state of the upper band empty, so that the ground state fulfills $\bra{G(Q)} \gamma^{\dagger}_{k-Q}\gamma_{k-Q} \ket{G(Q)} = 0$. In contrast, if $k \in S_{-}$, the lowest energy state is realized by occupying the upper band, so that the ground state fulfills $\bra{G(Q)} \gamma^{\dagger}_{k-Q}\gamma_{k-Q} \ket{G(Q)} = 1$.
The ground state $|G(Q)\rangle$ is therefore characterized by the following conditions
\begin{equation}
\begin{array}{lc}
 \gamma_{k-Q}|G(Q) \rangle=0 & \mbox{if }   k\in S_{+}\\
    \gamma^\dagger_{k-Q}|G(Q)\rangle=0 &   \mbox{if }    k\in S_{-} 
\end{array}\label{conditions_GS}
\end{equation}
and its energy is given by
\begin{equation}
		E_{0}\left(Q\right) = -\frac{1}{2} \sum_{k} \left|E_{+}(k;Q)\right| \leq 0 \quad,
		\label{GS-energy}
	\end{equation}
as follows from Eq.(\ref{diagonal}). The conditions (\ref{conditions_GS}) straightforwardly imply that 
\begin{equation}
\label{G-pre}
\ket{G(Q)} = \mathcal{N}\prod_{k \in S_{+}}\gamma_{k-Q }\prod_{k \in S_{-}}\gamma^{\dagger}_{k-Q }\ket{R} 
\end{equation}
where   $\mathcal{N}$ is a normalization constant and $\ket{R}$ is a reference state to be determined. Equation (\ref{G-pre}) is the general expression of ground state of the Kitaev model in the presence of superconducting modulation.  Depending on the  $k$-dependence of the spectrum, one can identify two different  parameter   regimes. 
 
 {\it  Gapped superconductor regime.} The first regime is characterized by a finite energy gap separation between the two bands. This occurs when $E_{+}(k;Q) > 0, \, \forall k \in [-\pi,\pi[$ or, equivalently, when $E_{-}(k;Q) < 0, \, \forall k \in [-\pi,\pi[$. 
  In this case the $S_{-}$ sector in Eq.(\ref{first_partition}) is trivially empty, and  the $S_{+}$ sector coincides with the entire BZ. The ground state (\ref{G-pre}) can then be written as $\ket{G(Q)} = \mathcal{N}\prod_{0<k<\pi}\gamma^{}_{k-Q }\gamma^{}_{-k-Q }\ket{R}$, the reference state $|R\rangle$ can easily be shown to coincide with the  electron vacuum,    $|R\rangle=|0\rangle$,     so that  $\ket{G(Q)}$  acquires the customary  expression  \begin{equation}
		\ket{G(Q)} = \!\prod_{0<k<\pi}\!\! \left(u_Q(k) + v_Q^{*}\left(k\right)c^{\dagger}_{-k-Q} c^{\dagger}_{k-Q}\right) \ket{0}\quad,
		\label{full_pairing_GS}
	\end{equation}
consisting of Cooper pairs in the bulk only. 
It can be shown  (see App.\ref{AppA}) that this regime exists for the following three parameter ranges
\begin{equation}\label{gapped-regime}
    \begin{array}{lcl}
      i) &  |\mu|>2w  \mbox{ \& } & \forall \Delta_0>0 \mbox{ \& } \forall Q\\ 
      ii) & |\mu|<2w  \mbox{ \&} & \sqrt{w^2-\mu^2/4}<\Delta_0 \,\mbox{ \& }   |\cos{Q}| \neq |\mu|/2w\\ 
      iii) & |\mu|<2w \mbox{ \& } & w|\sin{Q}|<\Delta_0<\sqrt{w^2-\mu^2/4} 
       \end{array}
\end{equation}
The case $\Delta_0>w$ discussed in \cite{ma-song2023, yanase2022} is a subcase of the second parameter range.\\ 

  {\it  Gapless superconductor regime.} 
 This  regime   occurs when 
 $E_{+}(k;Q)< 0$ (or equivalently $E_{-}(-k;Q) > 0$) for some values of $k$, i.e. when  the $S_{-}$ sector in Eq.(\ref{first_partition}) is not empty.  Recalling the expression (\ref{bands}), the condition for such gapless superconducting regime to occur are determined by imposing that the condition $|h_0(k;Q)|>h(k;Q)$ is fulfilled for some $k$'s. Some lengthy but straightforward algebra [see Appendix \ref{AppA}] leads to conclude that such a situation occurs if and only if the  following  parameter conditions are  both fulfilled
\begin{equation}\label{gapless-regime}
\left\{ \begin{array}{l}
\sqrt{\Delta_0^2+\frac{\mu^2}{4}} < w  \\
 \Delta_0   < w |\sin{Q}|  
\end{array}\right.\quad.
\end{equation}
Because  now the $S_{-}$ sector is not empty, the $\gamma^\dagger$ operators appearing in   the general Eq.(\ref{G-pre}) yield  some different features with respect to the conventional gapped superconductor, as can be suitably recognized by re-expressing Eq.(\ref{G-pre})
 in terms of fermionic operators $c^\dagger$. To this purpose, we introduce one further partitioning $S_{+}=S_{h} \cup S_{p}$ of the $S_{+}$ sector defined in Eq.(\ref{first_partition}),  where  the two subsectors $S_h$ and $S_p$ are identified by inspecting the sign of $E_{+}(-k)$ or, equivalently the sign of $E_{-}(k)$, in view of Eq.(\ref{rel_bands1}). Explicitly, one can partition the BZ as $S_{h} \cup S_e \cup S_{p}$, where
  \begin{equation}
		\begin{array}{ccl}
			k \in S_{h}  & \mbox{iff} &E_{\pm}(k;Q) > 0   \\
			k \in S_e & \mbox{iff} &   E_{\pm}(k;Q) < 0  \\
			k \in S_{p} & \mbox{iff} &   E_{+}(k;Q) >0\, \& \, E_{-}(k;Q) < 0 
		\end{array}\quad.
	\label{second_partition}
	\end{equation} 
Note that $S_{e}$ is just another notation for $S_{-}$.
In this case, one can show  [see Appendix~\ref{AppB}] that $\ket{R} = \prod_{k \in S_{h}} c^{\dagger}_{k-Q} \ket{0}$ and that $\ket{G(Q)}$ can be rewritten  in terms of the $c^{\dagger}_{k}$ operators as
	\begin{equation}
		\ket{G(Q)} =\!\! \prod_{\substack{0< k <\pi \\ k \in S_{p} }}\!\!\left(u_Q(k) + v^{*}_Q (k)c^{\dagger}_{-k-Q}c^{\dagger}_{k-Q}\right)\prod_{k \in S_e} c^{\dagger}_{k-Q}\ket{0}.
		\label{G-gen}
	\end{equation}
As compared to the gapped superconductor (\ref{full_pairing_GS}),  the gapless superconducting state (\ref{G-gen}) contains not only Cooper   pairs  ($S_{p}$ sector), but also   a pocket of \textit{unpaired electrons} ($S_e$ sector), and a pocket of \textit{unpaired holes} ($S_{h}$ sector). An illustrative example is shown in Fig.\ref{Fig2}(c), where the unpaired electron and hole pockets are highlighted in cyan and pink colors, respectively.  In this regime the superconducting order parameter is no longer interpreted as the gap. Yet, the energy (\ref{GS-energy}) of the state (\ref{G-gen}) is still lower than the state of a fully normal state ($\Delta_0\rightarrow 0$). 
Interestingly, the structure of Eq.(\ref{G-gen}) is similar to the one found for neutral superfluids with orbital angular momentum\cite{moroz2017,tada2018}. Here, however,  the $p$-wave nature of the order parameter implies significantly different features, as will be explained in Sec.\ref{sec4}.

As a consequence of its mixed structure, $\ket{G(Q)}$ exhibits normal and anomalous correlations depending on the $k$-sectors. Explicitly,  one can show [see Appendix~\ref{AppB}] that the normal correlation read
\begin{equation}
\langle c^\dagger_{k-Q} c^{}_{k'-Q}\rangle =\delta_{k,k'} \left\{ \begin{array}{lcl} 0 & \mbox{if} & k \in S_h \\
1 & \mbox{if} & k \in S_{e} \\
|v_Q(k)|^2 & \mbox{if} & k \in S_p
\end{array}\right. 
\end{equation}
\begin{equation}
\langle c^{}_{k-Q} c^{\dagger}_{k'-Q}\rangle =\delta_{k,k'} \left\{ \begin{array}{lcl} 1 & \mbox{if} & k \in S_h \\
0 & \mbox{if} & k \in S_{e} \\
u_Q^2(k) & \mbox{if} & k \in S_p
\end{array}\right. \label{corr-norm}
\end{equation}
while the anomalous correlations are
\begin{equation}
\langle c^{\dagger}_{k-Q} c^{\dagger}_{-k'-Q}\rangle =\delta_{k,k'} \left\{ \begin{array}{lcl} 0 & \mbox{if} & k \in S_h \\
0 & \mbox{if} & k \in S_{e} \\
-u_Q(k) v_Q(k) & \mbox{if} & k \in S_p
\end{array}\right. \,\,. \label{corr-anom}
\end{equation} 

The elementary excitations above the ground state $\ket{G(Q)}$ are given in Appendix \ref{AppC}. 
Before concluding this section, a remark is in order. In partitioning the BZ, we have not mentioned the case $E_{+}(k)=0$.  This corresponds to a degeneracy in the spectrum of the  Hamiltonian (\ref{diagonal}). Thus, although  one ground state can still be written in the forms  (\ref{full_pairing_GS}) or (\ref{G-gen}), it is degenerate with a state  $\ket{G^\prime(Q)}$ where an additional single fermion   is present. In a closed system, these two ground states have different fermion parity\cite{beenakker2013}. However, if the system is contacted to reservoirs inducing a current flow,  as is the case of interest here, fermion leakage makes both states equally probable. Thus, such a single fermion state merely represents here a zero-measure support set in the BZ.  \\

\section{Effective 2D fermion model and Lifshitz transition}
\label{sec4}
We now want to discuss how  the presence of the superconducting phase modulation $Q$ affects the topological aspects of the Kitaev model. 

When the system is in the gapped regime,   although $Q$ changes quantitatively the spectrum with respect to the $Q=0$ case [compare e.g. Figs.\ref{Fig2}(a) and (b)], it does not alter the {\it occupancy} of bands because $|h_0(k;Q)|<h(k;Q)\, \forall k$. The ground state $\ket{G(Q)}$ consists of a completely empty upper band~$E_{+}$ or, equivalently, of a completely filled lower band $E_{-}$. Then, one can characterize $\ket{G(Q)}$ through the topological index associated to the lower band.
Notice that  this case, described in Ref.\cite{ma-song2023}, follows exactly the same lines as the customary $Q=0$ case, and the role of $Q$ is to merely renormalize the hopping amplitude $w\rightarrow w \cos{Q}$ appearing in Eq.(\ref{xi}).

The situation is different in the   gapless regime, where $E_{+}(k;Q)<0$,  for some $k$'s. For such $k$-states the presence of the $h_0$-term in Eq.(\ref{bands}) alters the occupancy. In the ground state $\ket{G(Q)}$ not all the lower band states are occupied, and the approach adopted to topologically classify gapped phases cannot be straightforwardly applied to gapless phases \cite{kotetes2022}.   \\

However, the presence of the superconducting order parameter modulation $Q$ offers the opportunity to adopt a different perspective.
The idea  is that  $Q$ can be regarded as the wavevector of an extra synthetic dimension, in addition to $k$. In this way we can associate a 1D superconductor with superconducting phase modulation to a 2D fermionic model. The Lifshitz transitions\cite{volovik2017}  occurring in the topology of  the Fermi surface $\mathcal{F}$ of the half-filled 2D model enable one to characterize the ground state phases of the 1D superconductor.  In particular, we shall show  that the $p$-wave superconductor is associated to a 2D WSM, which can be in  type-I or type-II regime.

\subsection{Effective 2D fermion model}
\label{sec-2D}
We start by observing that, although the envisaged system is one-dimensional, the Hamiltonian (\ref{Ham-real-space}) is $2\pi$-periodic in the superconducting order parameter modulation $Q$. 
As a consequence,  the Bogolubov de Gennes Hamiltonian Eq.(\ref{Ham-BdG}) of the Kitaev chain with the superconducting phase modulation can also be interpreted as the first-quantized Hamiltonian of a two-dimensional system with a sublattice or orbital  degree of freedom $A/B$ 
\begin{equation}
\mathcal{H}_{2D}=\sum_{\bm{k}}\left( f^\dagger_{\bm{k}A},f^\dagger_{\bm{k}B}\right) {H}(\bm{k}) \left( \begin{array}{c}f^{}_{\bm{k}A}\\ f^{}_{\bm{k}B} \end{array}\right) \quad, \label{H2D}
\end{equation}
where $\bm{k}=(k,Q)$ is the wavevector lying on a torus.  In Appendix \ref{AppD}, we provide the explicit expression of $\mathcal{H}_{2D}$ in real space. 
The spectrum of this fictitious 2D model is determined by  ${H}(\bm{k})$ and is therefore the same as the one of the Kitaev model with a superconducting phase modulation. Two remarks are in order.

First, the 2D model contains twice the  degrees of freedom of the Kitaev model. Indeed, while in Eq.(\ref{H2D}) the spinors corresponds to two actual independent sublattice degrees of freedom and both bands   $E_{+}$ and $E_{-}$ are physical,  in the Kitaev model such two spectral bands are not independent in view of the redundancy intrinsic in the Nambu spinors, so that only one is actually physical.     

The second remark is related to symmetries. In such a dimensional promotion, the role of symmetries is interchanged. In particular,   while for the Kitaev chain the relations (\ref{time-reversal}) and (\ref{inversion})   encode {\it broken}   time-reversal   and inversion symmetries, for the associated  2D model  Eq.(\ref{H2D}) they represent    {\it fulfilled} symmetries. Recalling that $H(\bm{k})=h_0(\bm{k})\sigma_0+\bm{h}(\bm{k})\cdot \boldsymbol{\sigma}$, these symmetries imply
\begin{equation}\label{TRS}
    \mathcal{T}\, \mbox{\small symmetry} \rightarrow \left\{\begin{array}{lcl}
    h_0(\bm{k})&=&h_0(-\bm{k})\\
        h_1(\bm{k})&=&h_1(-\bm{k})\\
  h_2(\bm{k})&=&-h_2(-\bm{k})\\
   h_3(\bm{k})&=&h_3(-\bm{k})
    \end{array} \right. 
\end{equation}
and
\begin{equation}\label{IS}
    \mathcal{I}\, \mbox{\small symmetry} \rightarrow \left\{\begin{array}{lcl}
    h_0(\bm{k})&=&h_0(-\bm{k})\\
        h_1(\bm{k})&=&-h_1(-\bm{k})\\
  h_2(\bm{k})&=&-h_2(-\bm{k})\\
   h_3(\bm{k})&=&h_3(-\bm{k})
    \end{array} \right.\quad, 
\end{equation}
respectively, and yield the relation (\ref{rel_bands2}). In contrast, while for the 1D Kitaev chain Eq.(\ref{particle-hole}) is   a  built-in particle-hole symmetry stemming from the BdG formalism, it  represents an {\it anisotropy} constraint  for the 2D model, which implies
\begin{equation}\label{BdGS}
 \mbox{\small BdG constraint} \rightarrow \left\{\begin{array}{lcl}
    h_0(k,Q)&=&-h_0(-k,Q)\\
    h_1(k,Q)&=&-h_1(-k,Q)\\
    h_2(k,Q)&=&-h_2(-k,Q)\\
   h_3(k,Q)&=&h_3(-k,Q)
    \end{array} \right.\quad, 
\end{equation}
 and is responsible for the mutual relation (\ref{rel_bands1}) between the two bands.

\subsection{Lifshitz transition} \label{subsec:Lifshitz}
The Fermi surface $\mathcal{F}$ is defined   in the 2D BZ as the set of $\bm{k}=(k,Q)$ such that the eigenvalues fulfill $E_{+}(\bm{k})=0$ or $E_{-}(\bm{k})=0$. In particular, we shall inspect the possibility that the Fermi surface contains  nodes,   where the two    bands $E_{+}$ and $E_{-}$ touch, i.e. $E_{+}(\bm{k}_W)=E_{-}(\bm{k}_W)=0$. In view of Eq.(\ref{bands}), this equivalently corresponds to  the set of four equations $h_j(\bm{k}_W)=0$ ($j=0\ldots 3$) for two unknown coordinates of $\bm{k}_W$. From  symmetry arguments based on Eqs.(\ref{TRS}), (\ref{IS}) and the constraint  (\ref{BdGS}), one can deduce whether and how such nodes   occur. Indeed, the fact that both $\mathcal{T}$ and $\mathcal{I}$ symmetries hold implies $h_1(\bm{k})\equiv 0$, which eliminates one equation. Then,  from Eq.(\ref{BdGS}) and either (\ref{TRS}) or (\ref{IS})   we see  that $h_2$ must be odd in the $k$ component, and even in $Q$-component of $\bm{k}$. This is the case for the Kitaev model, because of the $p$-wave symmetry of the superconducting order parameter (\ref{delta}). This  means that the nodes can only occur at $k_W=0$ or $k_W=\pi$, and that they exist as long as $h_3=2w \cos{Q} \cos{k}-\mu$ [see Eq.(\ref{h-vec-def})]  can vanish, which is the case only if $|\mu|<2w$.  Then, the nodes are Weyl nodes  that are locally protected by $\mathcal{T}$ and $\mathcal{I}$ symmetries. Their    location
\begin{equation}\label{kW-def}
\begin{array}{lcl}
\bm{k}_W^{0,\pm}&=&(0,\pm Q_W) \\
\bm{k}_W^{\pi,\pm}&=&(\pi ,\pm (\pi-Q_W))
\end{array}\quad,
\end{equation}
where
     \begin{equation}\label{QW-def}
     Q_W(\mu) =\arccos(\mu/2w)\quad
     \end{equation} 
depends on the chemical potential $\mu$ and not on the value of $\Delta_0$. The WSM nature (Type-I {\it vs} Type-II) is  determined by the   behavior of the $h_0(\bm{k})$ term.
In the vicinity of the Weyl node $\bm{k}_W^{\lambda,\pm}$ (with $\lambda=0,\pi$), the Hamiltonian $H(\bm{k})$ in Eq.(\ref{Ham-BdG}) is well approximated by a low energy Hamiltonian 
\begin{equation}
H^{\lambda,\pm}\left(q_{2},q_{3}\right) = \alpha^{\lambda,\pm} q_{2}\sigma_{0} + \sum_{i,j=2}^3\, q_{i}\mathcal{V}^{\lambda,\pm}_{ij}\sigma_{j}, 
\end{equation}
where $q_{2}=k$ and $q_3=Q-Q^{\lambda,\pm}_W$ correspond to the deviations in momentum from $\bm{k}_W^{\lambda,\pm}$, and  
\begin{equation}\label{V-mat}
\begin{array}{lcl}
\mathcal{V}^{0,\pm}_{22} &=& -2\Delta_{0}\\\mathcal{V}^{0,\pm}_{33}  &=& -\alpha^{0,\pm}=\mp\sqrt{4w^{2} - \mu^{2}} \\ & & \\
\mathcal{V}^{\pi,\pm}_{22} &=&   2\Delta_{0} \\
\mathcal{V}^{\pi,\pm}_{33} &=& - \alpha^{\pi,\pm}=\pm \sqrt{4w^{2} - \mu^{2}}
\end{array}\quad.
\end{equation}
\\
Each Weyl node carries a vortex, as can be seen from  by computing   the lower band Berry phase  over a contour enclosing the Weyl node in the $\bm{k}$ momentum space 
\begin{equation}\label{Berry-phase}
\varphi^{\lambda,\pm} =\oint   d {\bf{k}} \cdot \bm{A}^{\lambda,\pm} = \pi\, \text{sgn}\left(\text{det}\mathcal{V}^{\lambda,\pm}\right) = \pm\pi, 
\end{equation}
 where $\bm{A}^{\lambda,\pm}$ is the Berry potential and $\mathcal{V}^{\lambda,\pm}$ is a diagonal matrix with components given by (\ref{V-mat}), near the Weyl  node   $\bm{k}_W^{\lambda,\pm}$.

Going back to the full BdG Hamiltonian (\ref{Ham-BdG})  and taking into account the explicit expressions (\ref{h0-def}) and (\ref{h-vec-def}), we deduce that for the Kitaev model   three scenarios can emerge in the Fermi surface of the associated 2D model, depending on the Kitaev model parameter ranges:\\

	\begin{figure*} 
		\includegraphics[scale = 0.7]{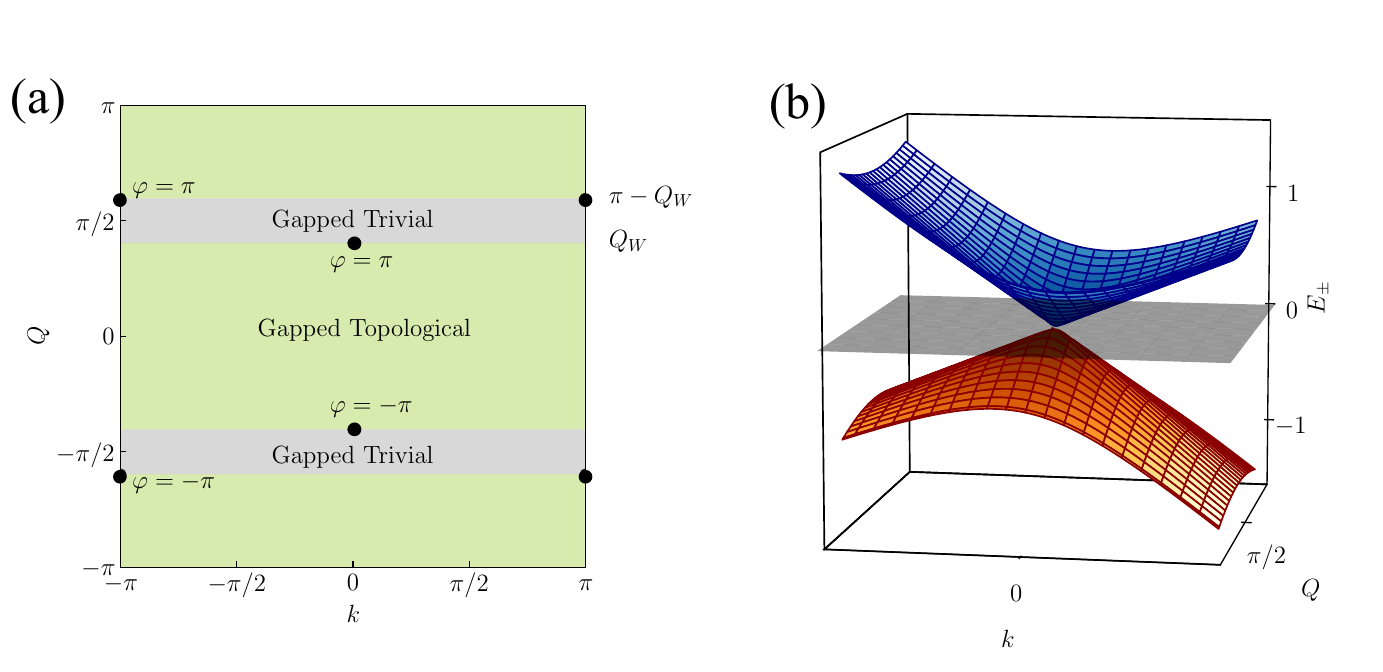}
  		\caption{\label{Fig3} The parameter range $2\sqrt{w^2-\Delta_0^2}  <|\mu|<2w$. (a) The Fermi surface of the 2D fermionic model associated to the Kitaev chain, with superconducting phase modulation $Q$ treated as a synthetic dimension. The Fermi surface consists of four Weyl nodes, highlighted with black bullets, located at positions given by Eq.(\ref{kW-def}). Here the plot is given for $\mu>0$. For $\mu<0$ the positions of $Q_W$ and $\pi-Q_W$ exchange. The green and gray areas denote the gapped topological and trivial phases of the 1D Kitaev chain, respectively. The value  of $Q_W$ determining the topological/trivial boundaries and the $Q$-coordinate of the Weyl nodes is given by Eq.(\ref{QW-def}) and depends only   on the chemical potential $\mu$ and not on $\Delta_0$.  (b) Energy band in the vicinity of one of the Weyl nodes, showing that the associated 2D fermionic model is a Type-I WSM. } 
	\end{figure*}

1)  {\it Type-I WSM phase.} In the parameter range 
\begin{equation} \label{Weyl-I-regime}
|\mu|<2w \hspace{0.2cm} \mbox{ \& } \hspace{0.2cm}  \sqrt{w^2-\mu^2/4}<\Delta_0 \,
\end{equation}
the Fermi surface   consists of four isolated  Weyl nodes~$\bm{k}_W$, where the two    bands $E_{+}$ and $E_{-}$ touch,  while $E_{+}>0$ and $E_{-}<0$ otherwise ($\bm{k}\neq \bm{k}_W)$. In this regime the $h_0$-term in the Hamiltonian modifies the spectrum but not the occupancy of the bands, and the model (\ref{H2D}) corresponds to a   2D  Type-I WSM with a Fermi level at $E=0$.  
The  Weyl nodes, shown as black bullets in Fig.\ref{Fig3}(a) and the type-I band dispersion relation is shown in Fig.\ref{Fig3}(b).

We shall now argue that this Fermi surface of the 2D model enables us to recover the topological classification of the Kitaev model and to determine the region of parameters where MQPs appear, in the presence of the superconducting phase modulation.  Indeed in this regime the lower band $E_{-}$ of the Hamiltonian ${H}$ is completely filled, while the upper band $E_{+}$ is completely empty. Thus, for each  cut of the band structure at  a fixed $Q$ away from the Weyl nodes ($Q\neq \pm Q^{0,\pi,\pm}_W$),  one   obtains a gapped one-dimensional  insulator, which can be classified in the  topological class  D\cite{schnyder2016}, since it only exhibits the particle-hole symmetry (\ref{particle-hole}).
The related  topological invariant is  the Zak-Berry phase of the lower band,  
 \begin{equation}\label{phi-ZB}
 \varphi_{ZB}=-i\int_{-\pi}^{\pi} \langle k,-|\partial_k |k,-\rangle \,dk\quad,
 \end{equation}
which is quantized in integer multiples of $\pi$ due to the particle-hole symmetry or, equivalently,  the index\cite{budich-ardonne2013}
\begin{eqnarray}
    \nu=(-1)^{ \varphi_{ZB}/\pi}&=&\mbox{sgn}\left(h_3(0) h_3(\pi)\right)= \nonumber\\
    &=& \mbox{sgn}\left(|\mu|-2w|\cos{Q}|\right)\quad.
\end{eqnarray}
Trivial phases ($\nu=+1$) correspond to the  condition $|\mu|> 2w |\cos{Q}|$,  whereas topological phases ($\nu=-1$) correspond to   $|\mu|< 2w |\cos{Q}|$ and, for the case $\mu>0$, are depicted as gray and green regions in Fig.\ref{Fig3}, respectively.
Thus, in the regime (\ref{Weyl-I-regime}) the range of $Q$-values where   MQPs exist is controlled by the chemical potential $\mu$, and is independent of the value of $\Delta_0$. The condition (\ref{Weyl-I-regime}) includes   the case $\Delta_0>w$ studied in Ref.\cite{ma-song2023}. In the case  $\Delta_0<w$,   Eq.(\ref{Weyl-I-regime}) reduces to
\begin{equation} \label{Weyl-I-regime-bis}
2\sqrt{w^2-\Delta_0^2}  <|\mu|<2w \quad,
\end{equation}
which, for the physically realistic regime $\Delta_0 \ll w$,   represents a very narrow  range of chemical potential value for this regime to exist. \\

\begin{figure*} 
		\includegraphics[scale = 0.7]{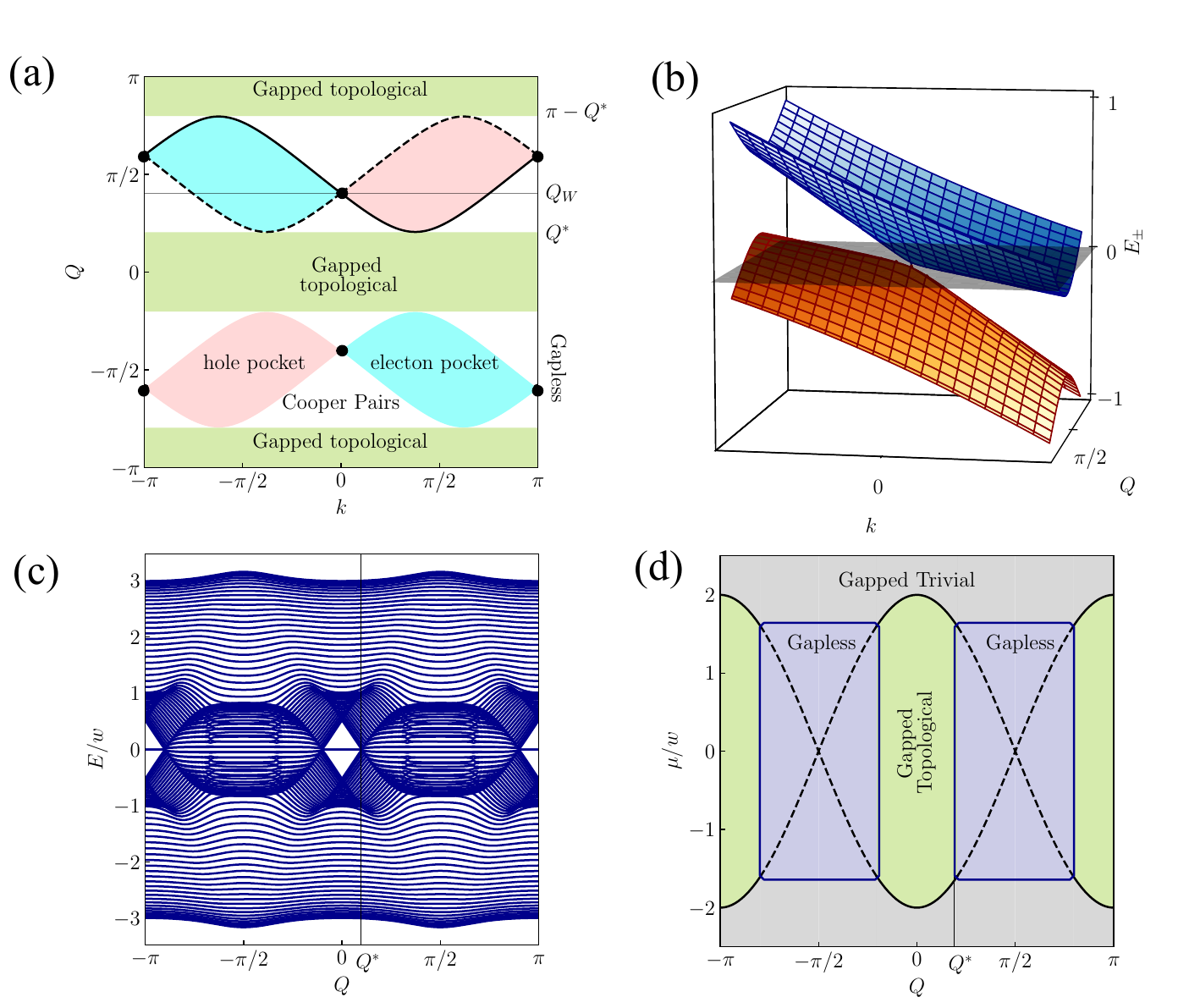}
  		\caption{\label{Fig4}   The parameter range $|\mu|<2\sqrt{w^2-\Delta_0^2}$. (a) The Fermi surface of the 2D fermionic model associated to the Kitaev chain with superconducting phase modulation $Q$ treated as a synthetic dimension. The Fermi surface consists of four Weyl nodes (black bullets), and of electron and hole pockets.  The green  areas denote the gapped topological  phases of the 1D Kitaev chain, while in the other the Kitaev model exhibits a gapless superconducting state, where both Cooper pairs and unpaired fermions and holes are present. The value  of $Q^*$ determining the topological gapped/gapless boundaries  only depends  on $\Delta_0$ [see  Eq.(\ref{Q-star-def})]. Solid and dashed lines enclosing the electron and hole pockets span the entire $k$-direction of the Brillouin zone, and belong to the homology class $(1,0)$. (b) energy band in the vicinity of one of the Weyl nodes, showing that the associated 2D fermionic model is a Type-II WSM.   (c)  spectrum of the Kitaev chain of $N_{s} = 50$ sites for $\mu = w$ and $\Delta_{0} = 0.3 w$. Zero energy Majorana modes can be seen in   the range $ |Q|< Q^* $ and $\pi- Q^*<|Q|<\pi$.  (d) $\mu$-$Q$ phase diagram of the system showing the gapped regions and the gapless regions, where solid and dashed black lines correspond to the band touching points, i.e. to the location of Type-I and Type-II Weyl nodes $Q_{W}\left(\mu\right)$, respectively [see Eq.(\ref{QW-def})].  }
	\end{figure*}

2) {\it Type-II WSM phase.}  
Let us now focus on the regime
\begin{equation}\label{Weyl-II-regime}
|\mu|<2\sqrt{w^2-\Delta_0^2} \quad,
\end{equation}
which  is    physically most realistic, as the superconducting order parameter is typically much smaller than the bandwidth parameter $w$. The scenario of the Fermi surface
  is depicted in Fig.\ref{Fig4}(a). The   Weyl nodes are still located in the positions $\bm{k}^W_{j,\pm}$ given by Eqs.(\ref{kW-def}). However,   they are now Type-II nodes, as shown in Fig.\ref{Fig4}(b), since  the Fermi surface of the  2D model  exhibits pockets of unpaired electrons and holes, which correspond to the regions where   the  magnitude of the $h_0(\bm{k})$-term overcomes the $h(\bm{k})$-term.  Note that electron and hole pockets are mutually mirrors of each other for $k\rightarrow -k$, as a straightforward consequence of the relation (\ref{rel_bands1}) originating from the  particle-hole constraint Eq.(\ref{particle-hole}) of the BdG formalism.
The three panels of Fig.\ref{Fig2} represent three cuts  in the 2D BZ shown in Fig.\ref{Fig4}(a), at   different values of $Q$, where one can see     the Lifshitz transition from the gapped phase  in panels (a)-(b) (empty Fermi surface)    to the gapless   phase in panel (c) (appearance of electron and hole pockets), where   the ground state of the Kitaev model is given by  Eq.(\ref{G-gen}).\\

The   fermion and hole pockets, identified by $E_{+}(\bm{k})=0$ and $E_{-}(\bm{k})=0$, respectively, cross at the Weyl nodes.  
 As one can see from Fig.\ref{Fig4}(b), in the range $Q^* < |Q|<  \pi-Q^*$, where 
\begin{equation}
Q^*(\Delta_0) = \arcsin\left(  \Delta_0/w  \right)\label{Q-star-def}
\end{equation}
the ground state is  gapless, and it exhibits Cooper pairs, unpaired fermions and holes. In contrast, in the range $ |Q|< Q^* $ and $\pi- Q^*<|Q|<\pi$,  it is  gapped and topological. Indeed the spectrum  plotted in Fig.\ref{Fig4}(c)  for a finite-size Kitaev chain of $N_s=50$ sites, shows the existence of zero-energy MQPs  in such $Q$ range of values.

A striking difference emerges   with respect to the regime 1) described above. In that case the range of $Q$-values, where the gapped topological phase exists is purely determined by the chemical potential $\mu$ and is {\it independent} of the superconducting order parameter $\Delta_0$, provided the latter is large enough to  fulfill the energy range (\ref{Weyl-I-regime}). This holds, in particular, for any $\Delta_0>w$. Indeed the topological phase --green region in Figs.\ref{Fig3}(a)-- is   determined by the $\mu$-dependent location of the Weyl nodes only.  In contrast, for $\Delta_0<w$, and in particular in the regime 2) specified by Eq.(\ref{Weyl-II-regime}), the $Q$-boundaries for the topological phases are determined by $\Delta_0$ through the critical value (\ref{Q-star-def}), and are {\it independent} of the chemical potential. As shown in Fig.\ref{Fig4}(a), the topological phase (green region) is no longer determined by the location of the Weyl nodes, but by the boundaries of the electron and hole pockets.
The comparison between the two regimes is clearly illustrated in Fig.\ref{Fig4}(d), where the   phase diagram of the Kitaev model in the regime 2) is shown  as a function of $\mu$ and $Q$, while the dashed lines represent the phase diagram for the regime 1). At each fixed value of $\mu$, the topological region  (green area) is smaller than in regime 1) because   the system enters the gapless phase (violet color).

3) {\it no Fermi surface.} When $|\mu| \rightarrow 2w$, the two  Weyl nodes $\bm{k}_W^{j,\pm}$ of each $j$-th pair  merge. Then,  in the regime
\begin{equation}\label{noFS}
|\mu|>2w \hspace{1cm} \forall \Delta_0>0
\end{equation}
the Weyl node equation $h_3=0$ cannot be fulfilled [see Eqs.(\ref{h-vec-def}) and (\ref{xi})], the Weyl nodes disappear and the Fermi surface is an empty set. This corresponds to the situation where $E_{+}(\bm{k})>0\,\, \forall \bm{k}$ and, in terms of a $Q$-cut, this describes a topologically trivial phase.\\

\subsection{Zero energy modes}
In open boundary conditions for the 2D model, both regimes 1) and 2) exhibit localized Fermi ``arc" states, which form effectively one-dimensional zero-energy bands, parametrized by the crystal momentum $Q$ along the infinite direction. We will consider for clarity a single boundary at $x=0$.

Let us start from the Type-I Weyl phase, in the parameter regime \eqref{Weyl-I-regime}. For $0<\mu<2w$, we find an arc connecting the projections of the nodes $\bm{k}_W^{0,\pm}$ and containing the point $Q=0$ and an arc connecting the projections of the nodes $\bm{k}_W^{\pi,\pm}$ containing $Q=\pi$, see Fig.\ref{Fig3}. Conversely, for $-2w<\mu<0$, the projection of the nodes $\bm{k}_W^{\pi,\pm}$ is closer to the point $Q=0$ than the projection of the nodes $\bm{k}_W^{0,\pm}$: in this situation, there is an arc connecting $\bm{k}_W^{0,\pm}$ across the BZ and an arc connecting $\bm{k}_W^{\pi,\pm}$ including the origin. In  terms  of the 1D model, the endpoints signal the transition between the topological and the trivial phase~\footnote{Note that in the Weyl-{I} regime (\ref{Weyl-I-regime}), if the denominator of $\lambda_{Q}$ is positive (topological phase), the argument of the square root in the numerator is also  positive.}.
In Appendix \ref{AppE}, we follow an established route for constructing Fermi arcs in Weyl  Hamiltonians \cite{Witten2016,Buccheri2022,Burrello2019,shen2013}. 
Linearizing the Hamiltonian \eqref{H2D} in $k$, one finds localized eigenstates in the form 
\begin{equation}\label{Fermi-arc}
\psi_Q(x,y)= \chi_Q(x) \, \frac{e^{i Q y}}{\sqrt{2 \pi}}
\end{equation}
where 
\begin{equation}\label{psi0}
     \chi_{Q} (x)=\frac{1}{\sqrt{\lambda_Q}}\left(\begin{array}{c}
e^{-i\zeta(Q)/2}\\
e^{i\zeta(Q)/2}
\end{array}\right)e^{-x/\lambda_Q}\quad.
\end{equation}
Here,  $\zeta(Q)=\arcsin{(w\sin Q/\Delta_{0})}$, while 
\begin{equation}\label{lambda}
\lambda_Q=
\frac{\sqrt{\Delta_{0}^{2}-w^{2}\sin^{2}Q}}{w \left|\cos Q\right|-\frac{\left|\mu\right|}{2}}
\end{equation}
represents the penetration depth. It is readily observed that this quantity {\it diverges} at the arc ends.

In the Type-II Weyl phase \eqref{Weyl-II-regime}, we see from Fig.\ref{Fig4} the emergence of electron/hole pockets in the bulk.   We find that the Fermi arcs are still described by \eqref{Fermi-arc}, but now connect the projections of the bulk electron and hole pockets onto the surface BZ. 
In particular,  the edge states have support between the values $\pm Q^*$ given in \eqref{Q-star-def}, including the origin $Q=0$, and outside the points $\pm \left(\pi-Q^*\right)$  [see Appendix \ref{AppE}]. In  terms  of the 1D model, these points signal the transition between the topological gapped phase and the gapless superconductor phase, see Fig.\ref{Fig4}. In contrast to the previous regime, the penetration length (\ref{lambda})  {\it vanishes} at the arc ends. A derivation and a plot of \eqref{lambda} in the two regimes can be found in Appendix \ref{AppE}.

The two components of the wavefunction \eqref{psi0} represent in the 2D model the weights on the $A$ and $B$ sublattices or orbital degrees of freedom. However, they are directly interpreted as the electron (top) and hole (bottom) components in the 1D superconductor model. Therefore, the wavefunction \eqref{psi0}   describes the Majorana edge mode of the 1D TS associated to the Fermi  arc  (\ref{Fermi-arc})   for the  Kitaev chain  with a $Q$-modulation of its superconducting order parameter. 
We recall that the 2D fermionic system (\ref{H2D}) exhibits twice the number of degrees of freedom as the original Kitaev model, for which  only a half of the spectrum (say, the upper band) can be retained. This is particularly evident in a slab geometry with a   finite size   along the $x$-direction. In the 2D fermionic model, two fermionic states  appear around $E=0$, and can be seen as symmetric and antisymmetric combinations of the Fermi ``arcs"  localized at the two edges. However, in the 1D Kitaev model   only {\it one} fermionic state   is physical  and it consists of {\it two} Majorana quasi-particles localized at the system edges.

\subsection{Comparison with the case of a $s$-wave superconductor}
Also a conventional $s$-wave superconductor carrying a current exhibits a modulation $Q$ of the superconducting order parameter, and can enter a gapless phase. Here we want to highlight the difference from the $p$-wave superconductor from the point of view of the mapping to a 2D model.

\begin{figure*}[ht]	
	\includegraphics[scale = 0.7]{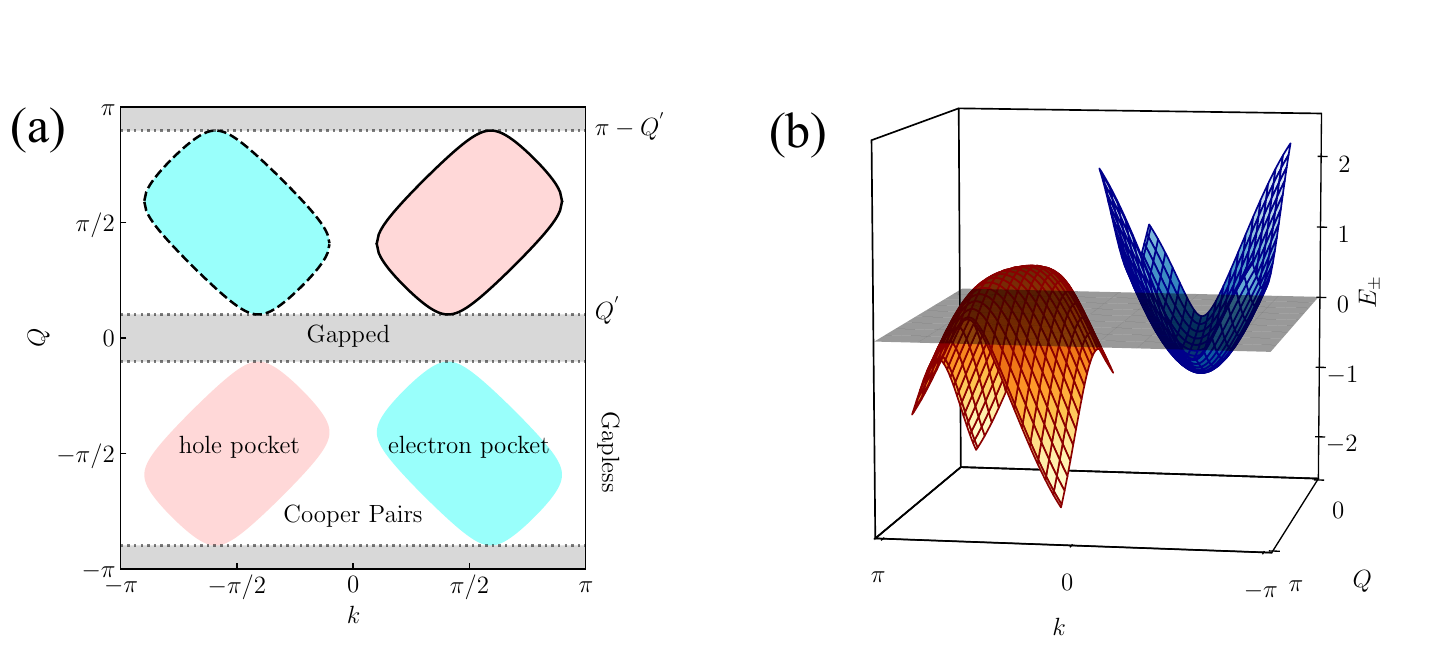}
	\caption{\label{Fig5} The case of an $s$-wave superconductor. $(a)$ Fermi surface for the 2D BCS model where the superconducting phase modulation $Q$ is treated as a synthetic dimension. The Fermi surfaces surrounding the electron (cyan) and hole (pink) pockets do not extend along the entire Brillouin zone rather remains localized. The gray areas denote  the gapped regions and determined by the boundaries $Q'$ and $\pi - Q'$ while the remaining regions show the coexistence of Cooper pairs and unpaired electrons and holes.  These pockets never touch each other, differently from what happens in the Kitaev model. As a consequence, solid and dashed lines enclosing the electron and hole pockets are localized over a finite portion of the   $k$-direction of the Brillouin zone, and belong to the homology class $(0,0)$.  $(b)$ Upper (blue) and lower (red) bands of the BCS model, in this case the associated 2D fermionic model corresponds to a conventional semimetal. 
 }
\end{figure*}

In a $s$-wave superconductor the  order parameter term  in the presence of a spatial modulation is $\Delta_{0}\sum_{k} (   c^{\dagger}_{k-Q \uparrow}c^{\dagger}_{-k-Q \downarrow} + {\rm H.c.} )$. By introducing a Nambu spinor 
$\Psi_{k;Q} = 
   (c_{k-Q \uparrow}, 
    c^{\dagger}_{-k-Q\downarrow}, 
    i\, c_{k-Q\downarrow},
   -i\, c^{\dagger}_{-k-Q \uparrow})$, the BCS Hamiltonian appears to be the sum $\mathcal{H}_{BCS} =(1/2) \sum_{k}\Psi^{\dagger}_{k;Q} (\sigma_0 \otimes {H}_{BCS})\Psi_{k;Q}$ of two decoupled sectors sharing the same BdG Hamiltonian ${H}_{BCS}(\bm{k})=h_0(\bm{k})\sigma_0+\bm{h}^\prime(\bm{k})\cdot \boldsymbol\sigma$, where $h_0$ is again given by Eq.(\ref{h0-def}), while
   \begin{equation}\label{h-vec-BCS-def}
  \bm{h}^\prime (k; Q)  =   \left(\Delta_0,\, 0 ,\, \xi(k;Q)\right) 
  \end{equation}
  with $\xi(k;Q)$ still given by Eq.(\ref{xi}). Similarly to what  described in Sec.\ref{sec-2D} for the Kitaev model, one can apply a mapping to a 2D (spinful) fermionic model, where spin is a degeneracy degree of freedom for the two bands. The  resulting Fermi surface, shown in Fig.\ref{Fig5}(a), exhibits two crucial differences with respect to the Kitaev case [see Fig.\ref{Fig4}(a)]. First,  the BCS model does not exhibit any node, since the   vector $\bm{h}^\prime$ in Eq.(\ref{h-vec-BCS-def}) can never vanish, as is clear from Eq.(\ref{h-vec-BCS-def}). As a related consequence,  gapped phases are always topologically trivial, since the Bloch vector $\hat{\bm{n}}=\bm{h}^\prime/|\bm{h}^\prime|$ spans a trivial line in the $h_1$-$h_3$ plane,  as expected for the BCS model. 
  
 Similarly to the Kitaev case, the presence of the $h_0$ term stemming from the superconducting modulation $Q$ leads to the emergence of gapless superconducting phases, which in this case  exist when both the conditions $Q^\prime<|Q|<\pi-Q^\prime$  and $Q^\prime<|k|<\pi-Q^\prime$ are fulfilled, where $Q^\prime=\arcsin(\Delta_0/2\sqrt{w^2-(\mu/2)^2})$. Note that these boundaries are dependent on {\it both} $\Delta_0$ and $\mu$, while in the Kitaev case they are $\mu$-independent. Importantly,   the boundaries of the electron and hole pockets  in the BCS case  are topologically different from the ones of   Fig.\ref{Fig4}(a). Indeed, while in the Kitaev model the solid and dashed boundaries run over the {\it entire} $k$-circle of the BZ, in the BCS model  they remain localized to the finite portion specified above.

Such a topological difference in the Fermi surfaces of $s$-wave and $p$-wave gapless superconductors can be characterized in terms of the homology group $H_1(T^2)$ of 1-cycles\cite{nakahara-book}, i.e. closed circuits on the 2D BZ torus $T^2$.  A closed Fermi surface circuit    is identified by a mapping $\lambda \in [-\pi,\pi] \rightarrow \bm{k}(\lambda)$ and can be classified according to   two topological indices $(n_k,n_Q)$, defined as the winding numbers related to the two orthogonal circuit direction along the torus 
 \begin{eqnarray}
     n_k&=&\frac{1}{2\pi}\int_{-\pi}^{\pi} \frac{dk}{d\lambda}\, d\lambda \\
          n_Q&=&\frac{1}{2\pi}\int_{-\pi}^{\pi} \frac{dQ}{d\lambda}\, d\lambda \quad.
 \end{eqnarray}
Let us now compare the Fermi surface closed circuits of the $p$-wave superconductor and $s$-wave superconductor. In the former case,     $\mathcal{F}$  consists of the four  curves that in  Fig.\ref{Fig4}(a) delimit  the electron and hole pockets. In particular, the ones highlighted as solid or dashed are given by
\begin{equation}
\left\{ \begin{array}{lcl}
k(\lambda)&=& \lambda\\
Q^\pm(\lambda)&=&\arccos\left(\frac{\mu}{2}\cos{k}\pm \sin{k} \sqrt{ 1-\frac{\Delta_0^2}{w^2} -\frac{\mu^2}{4w^2}  }\right)
\end{array}\right.
\end{equation}
respectively. Note that, at the   crossing at the Weyl nodes,  the solid and the dashed line are uniquely identified through the continuity of the eigenvectors related to the vanishing eigenvalues $E_{+}=E_{-}=0$. As a consequence, the Fermi surface in the Kitaev  case winds around the $k$-direction and therefore belongs to the $(1,0)$ topological class. In contrast, for the $s$-wave superconductor described by the BCS model, the  Fermi surface illustrated in Fig.\ref{Fig5}(a)  is always localized and belongs to the $(0,0)$ class, which is homotopically equivalent to a point. 

Importantly, these  different topological classes are 
 closely related  to the presence or absence of nodes in the 2D Fermi surface. Indeed in the Kitaev case the electron and hole pocket must necessarily go through the Weyl nodes, located at $k=0$ and $k=\pm \pi$, because the $p$-wave superconducting order parameter   is odd in $k$. In a $s$-wave superconductor, instead, the absence of nodes makes the Fermi surface consist of localized curves. 
 
Thus, in terms of the associated the 2D model,  a 1D $s$-wave superconductor corresponds to either a 2D insulator or a 2D conventional semimetal, depending on the parameter range [see Fig.\ref{Fig5}(b)]. In contrast, a 1D $p$-wave superconductor corresponds to a trivial 2D insulator or a 2D Type-I/Type-II WSM  [see Fig.\ref{Fig3}(a) and Fig.\ref{Fig4}(a)].

\section{$Q$-dependence of the current}
\label{sec5}
\begin{figure}[h]
 \centering
 \includegraphics[scale = 0.54]{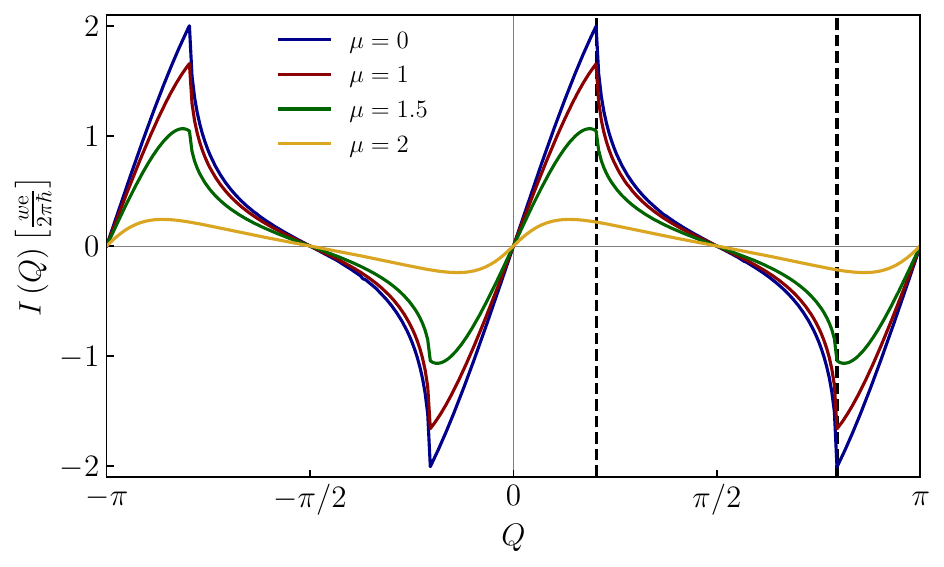}
 \caption{The current $I(Q)$ is plotted as a function of $Q$ for different values of $\mu$ and $\Delta_{0} = 0.6w$ showing the cusp signature of the Lifshitz transition. The vertical black dashed lines correspond to $Q^{*}$ and $\pi - Q^{*}$, where $Q^*$ is given by Eq.(\ref{Q-star-def}).}
 \label{Fig6}
 \end{figure}

We now want to analyze the behavior of the current carried by the ground state $\ket{G(Q)}$. As is well known, the operator describing the current flowing through the site~$j$ is given  by $
\hat{J}_j=- iw {\rm e} \, (  c^\dagger_{j}  c^{}_{j+1} - \, c^\dagger_{j+1}  c^{}_{j} )/\hbar
$, where~$-{\rm e}=-|{\rm e}|$ is the electron charge. The spatial conservation of the   expectation values $I(Q)=\langle G(Q)|\hat{J}_j|G(Q)\rangle$\,\, $\forall j$ follows from  charge conservation, and one can show  \footnote{By exploiting the current conservation at stationarity, one can write that $\langle {J}_j\rangle={N}_s^{-1} \langle \sum_j {J}_j\rangle$. Then, by applying the gauge transformation $c_j=e^{-i Q j} \tilde{c}_j$, one can shift the phase modulation in the superconducting term and set it into the hopping term $w\sum_j (e^{-i Q}\tilde{c}^\dagger_j \tilde{c}^{}_{j+1}+{\rm H.c.})$, whence one has  $\langle {J}_j\rangle=({\rm e}/{N}_s \hbar)\partial \langle \mathcal{H}\rangle/\partial{Q}$, and therefore Eq.(\ref{I(Q)}).}
 that
\begin{equation}
I(Q) =  \frac{{\rm e}}{\hbar} \frac{1}{N_s}  \frac{\partial E_{0} (Q)}{\partial Q} \quad. \label{I(Q)}
\end{equation}
Furthermore, exploiting the expression (\ref{GS-energy}) and taking the thermodynamic limit $N_{s}\rightarrow \infty$, one obtains the current  as a function of the superconducting modulation $Q$
\begin{eqnarray}
\lefteqn{I(Q) = -\frac{ w\, {\rm e}}{2\pi\hbar}     \int_{-\pi}^{\pi}dk 
\sin(k-Q) \times } \label{current} & &\label{current_continuum-v0} \\
& & \times \left(    \eta(k;Q)   + \frac{\xi(k;Q)}{ \sqrt{\xi^{2}(k;Q) + \left|\Delta_{k}\right|^{2}  } }   (1-|   \eta(k;Q)|) \right)  \nonumber
\end{eqnarray}
where   
\begin{equation}
\eta(k;Q)=\frac{1}{2}\left( \mbox{sgn}E_+(k;Q) +\mbox{sgn}E_-(k;Q)\right) \label{eta-def}
\end{equation}
is odd function of $k$, called spectral asymmetry\cite{moroz2017}, which identifies the three sectors (\ref{second_partition}) as well as the relative magnitudes of   $h_0$ and $h$ in Eqs.(\ref{h0-def})-(\ref{h-vec-def}), namely
\begin{equation}
\begin{array}{lcccl}
k\in S_h & \leftrightarrow & \eta=+1 & \leftrightarrow & h_0>h \\
k\in S_e & \leftrightarrow & \eta=-1 & \leftrightarrow & h_0<-h \\
k\in S_p & \leftrightarrow & \eta=0 & \leftrightarrow & |h_0|<h \\
\end{array}\quad.\label{eta-sectors-rel}
\end{equation}
\\
Figure \ref{Fig6} shows the current in units of $w{\rm e}/2\pi\hbar$ as a function of $Q$, for different values of the chemical potential~$\mu$.  The golden curve  illustrates the case, where the  model parameters fulfill the condition (\ref{Weyl-I-regime-bis}), corresponding to the Type-I phase for the WSM.  The   current   is    a smooth function of $Q$. In contrast, the blue and red  curves in Fig.\ref{Fig6} refer to   cases where   the condition (\ref{Weyl-II-regime}) is fulfilled, corresponding to the Type-II phase for the associated WSM. As one can see, the current exhibits sharp cusps at $Q=\pm Q^*$ and $Q=\pm (\pi-Q^*)$ [see Eq.(\ref{Q-star-def})], which are the hallmark of the   transition from the topological gapped to the gapless superconducting phase, where unpaired fermions and holes appear in the ground state [see Fig.\ref{Fig4}].  The location of the cusp is {\it independent} of the value of the chemical potential.   Note also that the cusps  correspond to the maximal current values, whereas the current always vanishes at $Q=\pm\pi/2$, for any value of $\mu$ and $\Delta_0$. From a mathematical viewpoint, this stems from the fact that the the current (\ref{current}) is an odd function of the deviation $q=Q\mp \pi/2$. From a physical viewpoint, for $Q=\pi/2$ the renormalized hopping term  $w\cos{Q}$ vanishes, the bare band $\varepsilon=2w \cos{k} \cos{Q}-\mu\rightarrow -\mu$ becomes flat, and unpaired fermions cannot carry any current, while  the crystal momentum $-2Q$ of the Cooper pairs equals its opposite $+2Q$, implying also that the Cooper pair current vanishes.

\begin{figure}[h]
	    	\includegraphics[scale = 0.54]{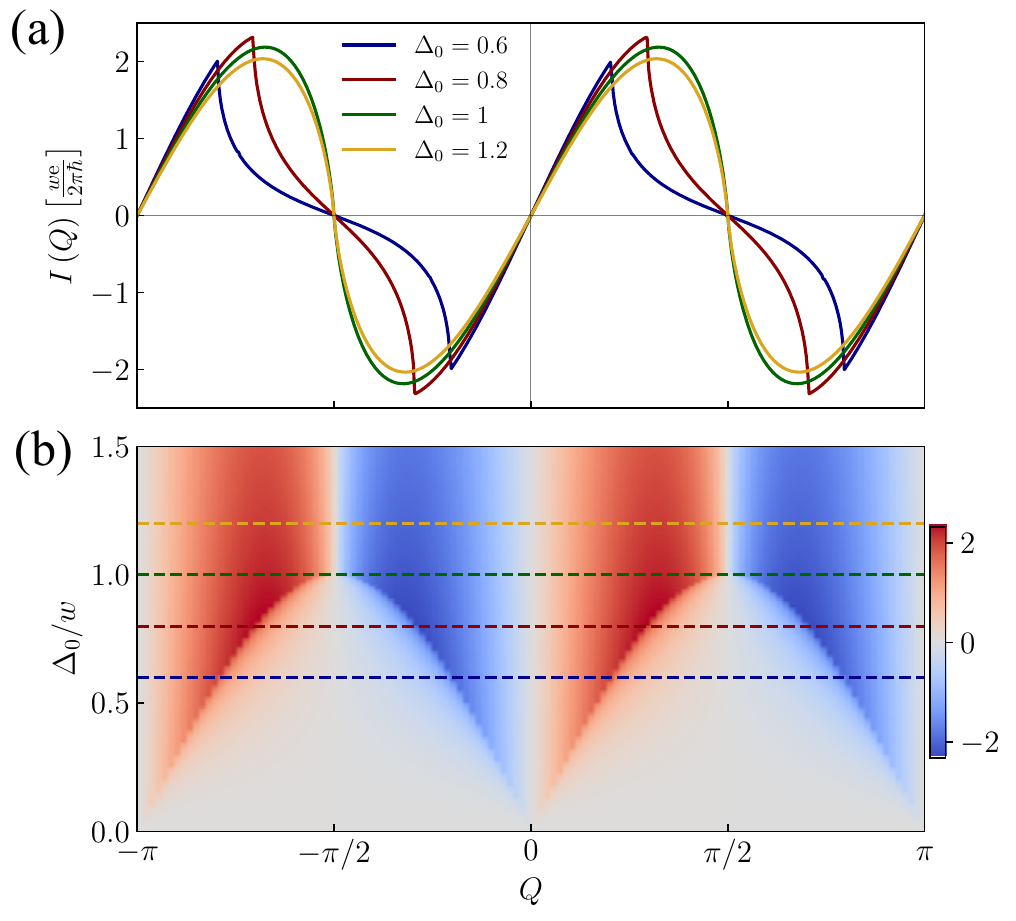}
	    	\caption{$\left(a\right)$ The current is plotted as a function of $Q$ for different values of $\Delta_{0}$ at $\mu = 0$. $\left(b\right)$ Contour plot of the current as function of $\Delta_{0}/w$ and $Q$ where the horizontal cuts (blue, red, gold and green dashed lines) correspond to the curves of panel $\left(a\right)$, the colors in the contour (red and blue) represent the current values (in units of $w {\rm e}/2\pi \hbar$).}
	\label{Fig7}
	    \end{figure}

	    The qualitatively different behavior of $I(Q)$ in the  Type-I {\it vs} Type-II parameter ranges (\ref{Weyl-I-regime-bis}) and (\ref{Weyl-II-regime}) is further  highlighted in Fig.\ref{Fig7}, which displays the current (\ref{I(Q)}) as a function of $Q$, at   $\mu = 0$, for different values   of the superconducting order parameter $\Delta_{0}/w$.  As one can see from panel $\left(a\right)$, for $\Delta_{0} > w$ the 1D Kitaev model is in the   WSM Type-I parameter range  and the current exhibits a smooth behavior as a function of $Q$, whereas for   $\Delta_{0} < w$ the Kitaev chain enters the WSM Type-II phase and the cusps clearly appear  in the current.   Panel (b) shows a density plot of the current as a function of $(Q,\Delta_0)$, where the horizontal dashed lines   represent the four cuts shown in panel (a).\\ 
For the special case $\mu=0$, it is also possible to determine an analytical expression for the current [see   Appendix \ref{AppF} for details], whence one can show that for $\Delta_0 \ll w$ the current exhibit a  linear behavior as a function of $Q$
\begin{equation}
    I(Q;\mu=0)\sim \frac{4  w {\rm e}}{2\pi \hbar}\, Q\quad,
\end{equation}
which represents the current carried by Cooper pairs with momentum $-2Q$ and charge $-2{\rm e}$. Similarly, for $\Delta_0 \ll w$,
  the maximal current $I^*$ reached at the cusp is
\begin{equation}\label{I-star}
    I^*(\mu=0)\sim \frac{4  \Delta_0 {\rm e}}{2\pi \hbar}\, 
\end{equation}
 For $\mu \neq 0$ one obtains a maximal current lower than the one at $\mu=0$, as can be seen from Fig.\ref{Fig6}. Thus, Eq.(\ref{I-star}) represents the maximal current that the TS, characterized by a given  $\Delta_0$,  can sustain in the topological phase, i.e. right at the transition to the gapless phase. \\

\begin{figure}[h]
		\centering
		\includegraphics[scale = 0.54]{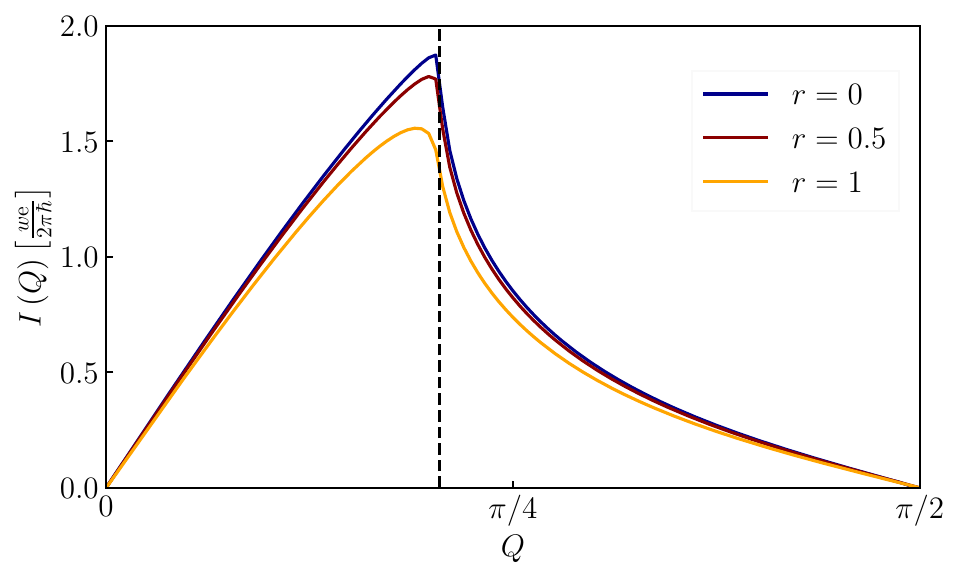}
		\caption{Effects of disorder: The current is shown as a function of $Q$ for parameters $\Delta_0=0.6w$, $\mu=0$ and system size $N_{s} = 80$, for three different values of on-site disorder strength, $r = 0$ (clean case), $r=0.5$ (moderate disorder), and $r=1$ (strong disorder).} 
		\label{Fig8}
	\end{figure}

{\it Effects of disorder.}
We have also analyzed the effects of disorder on the behavior of the current. Specifically, we have considered a disordered on-site potential, which we have  accounted for by replacing the constant chemical potential in  the Hamiltonian (\ref{Ham-real-space}) with $\mu\rightarrow \mu_{j} = \mu + w\,r\,\rho_{j}$, where $r$ is the  disorder strength parameter and $\rho_{j} \in \left[-1,1\right]$ is a uniformly distributed random number. Figure \ref{Fig8}  shows the $Q$-dependence of the current for $\Delta_{0} = 0.6$, $\mu = 0$, for three values of disorder strength, $r = 0$ (clean case), $r = 0.5$ (moderate disorder) and $r = 1$ (strong disorder). Each curve corresponds to an average over a large number of disorder realizations.  As far as the disorder remains moderate, the cusp  behavior of the current at $Q=Q^{*}$ (vertical dashed line) remains robust, and the only effect of disorder is to reduce a bit the magnitude of the current. Only for strong disorder the curve becomes smooth and the cusp disappears.     
\\
	
\section{Discussion and Conclusions}\label{sec6}
In this paper we have investigated the effects of the   spatial modulation $Q$ emerging in the superconducting order parameter when  a  topological $p$-wave  superconductor carries a  current flow.
By modelling the system with  a Kitaev chain, we have analyzed the properties of  the current carrying state  for arbitrary values of the model parameters, including the physically realistic regime $\Delta_0<w$. 

We have demonstrated that, by treating the $Q$-modulation as   the wavevector related of an extra synthetic dimension,  it is possible to establish a mapping between the stationary nonequilibrium state of the  1D superconductor and the ground state of a 2D WSM. The Lifshitz transitions emerging in the Fermi surface of the 2D model as a function of the model parameters identify different phases of the topological superconductor crossed by a current. Exploiting such a mapping, we have also  constructed the  expression (\ref{Fermi-arc}) of the Fermi ``arc"  in the 2D WSM and the Majorana quasi-particle (\ref{psi0}) in the corresponding current-biased 1D TS.

In particular, in the regime identified by Eq.(\ref{Weyl-I-regime}), the 2D WSM is in its Type-I phase with four Weyl nodes, whose location  depends only on the chemical potential $\mu$ through the value of $Q_W$ [see Eq.(\ref{QW-def})], and determines the separation   between gapped topological and trivial  phases of the Kitaev chain. The Fermi ``arcs"  (\ref{Fermi-arc}) correspond to the Majorana quasi-particles (\ref{psi0}) emerging in the topological phase (see Fig.\ref{Fig4}),  and their penetration length (\ref{lambda}) diverges at $Q_W$ and $\pi-Q_W$. Note that, for realistic values $\Delta_0 \ll w$, the regime  Eq.(\ref{Weyl-I-regime-bis}) of Type-I WSM reduces to a small chemical potential range. 
In contrast, when the model parameters fulfill the broader range Eq.(\ref{Weyl-II-regime}), the 2D system is a 
Type-II WSM. In this case  electron and hole pockets   correspond to the gapless superconductor phases of the Kitaev chain, where  Cooper pairs coexist with unpaired electrons and holes (see Fig.\ref{Fig5}).  In this case, the $Q$-range where the gapped topological phase exists is   no longer determined by the location of the Weyl nodes. Instead, it depends   on $\Delta_0$ only, through the value of $Q^*$ given by Eq.(\ref{Q-star-def}). Notably,  the penetration length of the Majorana edge mode vanishes at such  transition value between the topological gapped and the gapless phase of the Kitaev model. 

Furthermore, we have  highlighted the difference with respect to the $s$-wave superconductor, where the associated 2D fermionic model can be either a trivial insulator or a conventional semimetal (see Fig.\ref{Fig6}). The difference between the 2D Fermi surfaces corresponding to the $p$-wave and $s$-wave superconductors are shown to correspond to two different classes of the homology group of closed circuits along the 2D BZ.

Finally, by computing the current   flowing through a the Kitaev chain, we have shown in  Figs.\ref{Fig6} and \ref{Fig7} that it exhibits a sharp cusp at the critical value $Q^*$,  corresponding to the transition between the gapped and the gapless phase illustrated in Fig.\ref{Fig5}(a) and (d).  Importantly, the value of $Q^*$ given  in Eq.(\ref{Q-star-def}) is independent of the chemical potential variations and is robust to disorder (see Fig.\ref{Fig8}). Moreover, it determines the maximal current $I^*$ [see Eq.(\ref{I-star})] that the $p$-wave topological superconductor can sustain in its topological phase where   Majorana quasi-particles exist. Before concluding, we also would like to discuss some possible implementations of the model investigated here, and to outline some future perspectives of our results.

{\it Implementations.} One of the most explored  implementations of 1D topological superconductors are nanowires with strong spin-orbit coupling, such as InSb and InAs, proximized by a thin conventional superconducting layer (e.g. Al or Nb) and exposed to a longitudinal magnetic field. One of the signatures that seems to be compatible with the existence of MQPs is a pronounced zero-bias peak observed in the conductance  in electron tunneling experiments.  However, the question whether such a peak is actually due to MQPs or to other nontopological effects is still under debate.  
Here, from our result Eq.(\ref{I-star}), we can estimate the maximal current that the 1D $p$-wave superconductor can sustain before entering the gapless phase, where Majorana quasi-particles are absent. Using  the estimated values $\Delta_0\sim 0.09 \div 0.25\,{\rm meV}$ for  the induced gap in InSb nanowires\cite{kouwenhoven2012,furdyna2012,kouwenhoven2018} and $\Delta_0\sim 0.05\,{\rm meV}$ for InAs nanowires\cite{heiblum2012}, one finds $I^* \sim 8-40\,{\rm nA}$. This range of values is of the same order of magnitude as the typical current in nanowire experiments.\cite{marcus2017,song2022,aghaee2023}

{\it Future perspectives.} Our analysis has focussed on the the mapping between a 1D TS and a 2D WSM. However, the idea of harnessing the phase modulation $Q$ as a synthetic dimension is actually quite general and can apply to higher dimensional cases. We thus can expect that a 2D TS where a current flowing in a specific direction can be mapped to 3D WSM. Our results thus pave the way to interpret  the properties of WSMs as an observation of lower dimensional TSs with a current flow.

\acknowledgments
F.D. acknowledges financial support from the NEThEQS PRIN 2022 Project (code 2022B9P8LN) funded by the Italian Ministry of University and Research and fruitful discussions with Mario Trigiante. F.G.M.C. acknowledges financial support   from ICSC—Centro Nazionale di Ricerca in High-Performance Computing, Big Data, and Quantum Computing (Spoke 7), funded by European Union -- Next Generation EU (Grant No. CN00000013).

	\appendix

	\section{Parameter ranges of gapless and gapped phases of the Kitaev model}
 \label{AppA}
Here we provide details to determine  the range of  parameters  $\Delta_0$, $\mu$ and $Q$ characterizing the gapped and gapless phases of the Kitaev  model.  Recalling that the spectrum Eq.(\ref{bands}) is determined by the functions $h_0$ and $h$ [see Eqs.(\ref{h0-def}) and (\ref{h-vec-def}) in the Main Text], we shall identify here the gapless and gapped phases.

{\it Gapless phase}. The gapless phase of the Kitaev model   is determined by the condition 
\begin{equation}\label{ineq-1}
    |h_0(k;Q)|>h(k;Q) \hspace{1cm} \mbox{for some $k$ and $Q$}
\end{equation}  
Introducing the quantity 
\begin{eqnarray}
\lefteqn{\Phi(k;Q)= \frac{1}{4}\left(h^2-h_0^2\right)=} & & \label{Phi-def}\\
&=& X_{Q}^2  - \frac{\mu}{w}X_Q  \cos{k} +  \frac{\mu^2}{4w^2}-  \left(1-\frac{\Delta_0^2}{w^2}\right)\sin^2{k}   \hspace{0.5cm}\nonumber
\end{eqnarray}
where $X_Q=\cos{Q}$, and 
recalling the expressions (\ref{h0-def}) and (\ref{h-vec-def}), 
the inequality (\ref{ineq-1}) amounts to requiring
\begin{eqnarray}
\Phi(k;Q) \,<0   \hspace{1cm}  \mbox{for some $k$ and $Q$} \quad.
\label{ineq-2}
\end{eqnarray}
At any given $k$  the inequality Eq.(\ref{ineq-2}) is fulfilled only 
in the range
\begin{eqnarray}
 X_{Q}^{-}(k) \le X_Q \le X_{Q}^{+}(k)  \label{range_Xphi-pre-1}
\end{eqnarray}
provided that the two roots \begin{equation}
 X_{Q}^\pm(k)= \frac{\mu}{2w}  \cos{k}\pm |\sin{k}|  \sqrt{ 1-\frac{\Delta_0^2}{w^2} -\frac{\mu^2}{4w^2}  }  \label{Xpm}
\end{equation}
of the  equality $\Phi=0$
are real. 
Thus, the requirement $X^\pm_Q \in \mathbb{R}$ implies 
\begin{equation}
   \Delta_0^2  +\frac{\mu^2}{4}<w^2 \hspace{0.5cm} \Rightarrow \hspace{0.5cm} \frac{|\mu|}{2w} <1  \hspace{0.2cm},\hspace{0.2cm}  
 \frac{\Delta_0}{w} <1  
 \end{equation}
which is the first Eq.(\ref{gapless-regime}) given in the Main Text. Then, we observe that, by introducing the following angle  
\begin{equation}
\theta=\arcsin\left( \frac{\frac{\mu}{2w}}{\sqrt{1-\frac{\Delta_0^2}{w^2}}}\right) \,\in \left[-\frac{\pi}{2},+\frac{\pi}{2}\right] \label{theta-def}
\end{equation}
and  by denoting $\sigma_k=\mbox{sgn}(\sin{k})$, 
the roots (\ref{Xpm}) can be rewritten as 
\begin{eqnarray}
X^\pm_Q =  \sqrt{1-\frac{\Delta_0^2}{w^2}} \sin\left(\theta  \pm \sigma_k k  \right) \quad. \label{Xpm-ter}
\end{eqnarray}
This implies that $  |X^\pm_Q| \le \sqrt{1-\frac{\Delta_0^2}{w^2}} <1$, i.e. that the solutions
(\ref{range_Xphi-pre-1}) always belong to the 
  physically acceptable range for the variable   $X_Q \in [-1 ; 1]$.
 Thus, at any given~$k$,   the range of $Q$ values for which unpaired electrons  and holes exist is given by  
{\small
\begin{eqnarray}
-\sqrt{1-\frac{\Delta_0^2}{w^2}}\, \le X_{Q}^{-}(k) \le X_Q \le X_{Q}^{+}(k) \le \sqrt{1-\frac{\Delta_0^2}{w^2}}\hspace{0.4cm}.\label{range_Xphi-pre-2}
\end{eqnarray}  
}
Because in Eq.(\ref{range_Xphi-pre-2}) the extremal values $\pm \sqrt{1-\Delta_0^2/w^2}$ are always reached by $X_Q^\pm(k)$ for some value of $k \in [-\pi,\pi]$, the actual range of $Q$ for which unpaired electrons and holes are present is precisely
\begin{eqnarray}
-\sqrt{1-\frac{\Delta_0^2}{w^2}}\,  \le \cos Q \le  \, \sqrt{1-\frac{\Delta_0^2}{w^2}}\label{cond-param-2-pre}
\end{eqnarray}
i.e. $
   \Delta_0< w|\sin{Q}| $. This is the second Eq.(\ref{gapless-regime}) given in the Main Text. In turn, it also determines the onset value (\ref{Q-star-def}) for the appearance of the electron and hole pockets.\\

{\it Gapped phase.} The Kitaev model is gapped if 
\begin{equation}
    |h_0(k;Q)|<h(k;Q) \,\, \, \forall k \in [-\pi;\pi[,
\end{equation}
i.e. if
\begin{eqnarray}
\Phi(k;Q) \,>0   \hspace{1cm}  \,\, \, \forall k \in [-\pi;\pi[\quad.
\label{ineq-gapped}
\end{eqnarray}
where $\Phi(k;Q)$ is given in Eq.(\ref{Phi-def}).
One can now distinguish two ranges of chemical potential. For $|\mu|>2w$, it is straightforward to realize from Eq.(\ref{Phi-def}) that the gapped phase condition Eq.(\ref{ineq-gapped}) is fulfilled $\forall\, Q$. This is the case i) in Eq.(\ref{gapped-regime}). When $|\mu|<2w$, Eq.(\ref{ineq-gapped}) is fulfilled  in two subcases. The first one is when $\Phi$ has no real roots for any $k$, which occurs for the parameter range ii) given in Eq.(\ref{gapped-regime}). The second one occurs when $\Phi$ has two real roots, i.e. for $\Delta_0^2+(\mu/2w)^2<w^2$. In that case, for a given $k$,  the inequality $\Phi(k;Q)>0$  is fulfilled for $-1\le X_Q<X_Q^{-}(k)$ and $X_Q^{+}(k)<X_Q\le 1$, where the roots are given by Eq.(\ref{Xpm-ter}).  However,  Eq.(\ref{ineq-gapped}) requires that this holds for {\it any} $k$, this implies that $|X_Q|>\sqrt{1-\Delta^2_0/w^2}$, i.e.  $|\sin{Q}|<\Delta_0$, whence one obtains the case iii) given in Eq.(\ref{gapped-regime}).

\section{Ground State}
 \label{AppB}
Here we show that   the general expression (\ref{G-pre}) of the current carrying ground state implies the equivalent expression (\ref{G-gen}). Indeed, by rewriting Eq.(\ref{G-pre})  using the partitioning (\ref{second_partition}) one can write the ground state as
	\begin{equation}
		\ket{G} = \mathcal{N} \prod_{k \in S_{p}} \gamma_{k-Q} \prod_{k \in S_{h}}\gamma_{k-Q} \prod_{k \in S_e} \gamma^{\dagger}_{k-Q} \ket{R},
	\end{equation}
 By changing the $k \rightarrow -k$ in the $\gamma$'s of the $S_{h}$ sector one can equivalently write
 	\begin{eqnarray}
 		\ket{G} &=& \mathcal{N} \prod_{k \in S_{p}} \gamma_{k-Q} \prod_{k \in S_e} \gamma_{-k-Q}\gamma^{\dagger}_{k-Q} \ket{R} = \nonumber \\
   &=& \mathcal{N}\prod_{ \substack{ 0<k<\pi \\k \in S_{3} } } \gamma_{-k-Q} \gamma_{k-Q} \prod_{k \in S_{2}} c_{-k-Q}c^{\dagger}_{k-Q} \ket{R},
 		\label{gs_pre} 
 	\end{eqnarray}
 	where we have used the relation $
 		\gamma^{}_{-k-Q}\gamma^{\dagger}_{k-Q} =  c^{}_{-k-Q}c^{\dagger}_{k-Q}$ following from Eqs.(\ref{Bogoliubov-quasi}). From Eq.(\ref{gs_pre}) we deduce that  the reference state is $\ket{R} = \prod_{k \in S_e} c^{\dagger}_{-k-Q} \ket{0}=\prod_{k \in S_{h}} c^{\dagger}_{k-Q} \ket{0}$. Finally, by noticing that $
\gamma^{}_{-k-Q}\gamma^{}_{k-Q} \ket{0}  =v_k^* (u_{k} +v_{k}^* \, c^{\dagger}_{-k-Q}c^{\dagger}_{k-Q}  )\ket{0}$, one obtains the normalized state Eq.(\ref{G-gen}). \\

	Moreover, the general expression (\ref{G-pre}) of the ground state also leads to Eqs.(\ref{corr-norm})-(\ref{corr-anom}). Indeed  the relations (\ref{conditions_GS}) stemming from Eq.(\ref{G-pre}) imply that:\\
	for $k,k' \in S_{+}$ and $-k,-k' \in S_{+}$  $\rightarrow k,k^\prime \in S_{p}$
		\begin{equation}
			\begin{array}{ccc}
				\langle \gamma^{\dagger}_{k-Q}\gamma_{k'-Q} \rangle &=& 0\\
				\langle \gamma^{\dagger}_{k-Q}\gamma^{\dagger}_{-k'-Q} \rangle & = & 0 \\
				\langle \gamma_{-k-Q}\gamma_{k'-Q} \rangle &=& 0 \\
				\langle \gamma_{-k-Q}\gamma^{\dagger}_{-k'-Q} \rangle & = & \delta_{k,k'}
			\end{array}
		\end{equation}
		for $k,k' \in S_{+}$ and $-k,-k' \in S_{-}$  $\rightarrow k \in S_{h},\, -k \in S_e$ 
		\begin{equation}
			\begin{array}{ccc}
				\langle \gamma^{\dagger}_{k-Q}\gamma_{k'-Q} \rangle &=& 0\\
				\langle \gamma^{\dagger}_{k-Q}\gamma^{\dagger}_{-k'-Q} \rangle & = & 0 \\
				\langle \gamma_{-k-Q}\gamma_{k'-Q} \rangle &=& 0 \\
				\langle \gamma_{-k-Q}\gamma^{\dagger}_{-k'-Q} \rangle & = & 0
			\end{array}
		\end{equation}
		for $k,k' \in S_{-}$ and $-k,-k' \in S_{+}$  $\rightarrow k \in S_e,\,-k \in S_{h}$
		\begin{equation}
			\begin{array}{ccc}
				\langle \gamma^{\dagger}_{k-Q}\gamma_{k'-Q} \rangle &=& \delta_{k,k'}\\
				\langle \gamma^{\dagger}_{k-Q}\gamma^{\dagger}_{-k'-Q} \rangle & = & 0 \\
				\langle \gamma_{-k-Q}\gamma_{k'-Q} \rangle &=& 0 \\
				\langle \gamma_{-k-Q}\gamma^{\dagger}_{-k'-Q} \rangle & = & \delta_{k,k'}
			\end{array}
		\end{equation}	
By inverting the Bogoliubov quasi-particle relations (\ref{Bogoliubov-quasi}) 
one obtains
\begin{equation}\label{c-intermsof-gamma-old}
 \left\{ \begin{array}{lcl}
  c^{}_{k -Q} &= & u_{k}\gamma^{}_{k-Q}    -v^*_{k}\, \gamma^\dagger_{-k-Q}  \\	    c^{\dagger}_{k-Q} &= & u_{k}\gamma^{\dagger}_{k-Q}  -v_{k}\gamma^{}_{-k -Q}  
\end{array}  \right.   
\end{equation}
whence Eqs.(\ref{corr-norm})-(\ref{corr-anom}) are straightforwardly deduced.\\

	\section{Elementary Excitations}
 \label{AppC}
	From Eq.(\ref{diagonal}) and from the properties (\ref{first_partition}) characterizing the ground state, it is straightforward to realize that the elementary excitations of the system are  characterized by an energy $|E_+(k;Q)|$ with respect to the ground state energy (\ref{GS-energy}), and are given by $\gamma^\dagger_{k-Q}|G(Q)\rangle$ for $
 k \in S_{+}$, and $ \gamma^{}_{k-Q}|G(Q)\rangle$ for $ k \in S_{-}$.
 Moreover, as we have seen above,   the BZ can be partitioned in three sectors ${\rm BZ}=S_h \cup S_{e} \cup S_p$, and one can identify {\it three} types of elementary excitations
 \begin{eqnarray}
|ex_{1}(k;Q)\rangle=\gamma^\dagger_{k-Q}|G\rangle & & \mbox{ for }
 k \in S_{h}\\
|ex_{2}(k;Q)\rangle=\gamma^{}_{k-Q}|G\rangle & & \mbox{ for }
 k \in S_{e}\\
|ex_{3}(k;Q)\rangle=\gamma^{\dagger}_{k-Q}|G\rangle & & \mbox{ for }
 k \in S_{p}
\end{eqnarray}
	Using Eq.(\ref{Bogoliubov-quasi}) we can see that the first type of excitation ($k \in S_h$) takes the form  
\begin{eqnarray}
\lefteqn{|ex_1(k;Q)\rangle =   \prod_{{\tiny \begin{array}{c}   k'>0 \\ k' \in S_{p}\end{array}}} \!\!\! \left(u_{k'} +v_{k'}^* \, c^{\dagger}_{-k'-Q}c^{\dagger}_{k'-Q}  \right) \times} \nonumber \\ 
& & \times \left(\prod_{{\tiny \begin{array}{c} k' \in S_{e} \\k' \neq -k \end{array}}}\!\!\!\!  c^{\dagger}_{k'-Q} \right)  \,  \left( v_k- u_k c^\dagger_{-k-Q}c^\dagger_{k-Q}  \right) 
 \ket{0}   \label{ex-1}
\end{eqnarray}
where, as compared to the ground state (\ref{G-gen}), an empty fermionic state at $k\in S_h$ and a filled fermionic state at $-k \in S_e$ are replaced by a cooper pair $v_Q(k)- u_Q(k) c^\dagger_{-k-Q}c^\dagger_{k-Q}$.  Notably, such type of Cooper pair is {\it orthogonal} to the Cooper pair characterizing the ground state, i.e.
\begin{equation}
\langle 0 | \left( u_{k} +v_{k} \, c^{}_{-k-Q}c^{}_{k-Q}\right) | \left( v_k- u_k c^\dagger_{-k-Q}c^\dagger_{k-Q}  \right) 
 \ket{0}=0
\end{equation}
For the $S_e$ sector the excited state reads
\begin{eqnarray}
\lefteqn{|ex_2(k;Q)\rangle  =    \prod_{{\tiny \begin{array}{c}   k'>0 \\ k' \in S_{p}\end{array}}} \!\!\! (u_{k'} +v_{k'}^* \, c^{\dagger}_{-k'-Q}c^{\dagger}_{k'-Q}  )  \times } & & \nonumber \\
& & \times \left(\prod_{{\tiny \begin{array}{c} k' \in S_{f} \\k' \neq -k \end{array}}}\,  c^{\dagger}_{k'-Q} \right) \left( u_Q(k)+ v_Q^*(k) c^\dagger_{-k-Q}c^\dagger_{k-Q}  \right) 
 \ket{0}  \label{ex-2}
\end{eqnarray}
where, as compared to the ground state (\ref{G-gen}), a single fermionic state at $k\in S_{e}$ and an empty fermionic state at $-k \in S_h$ are replaced by a cooper pair $u_k+ v_k^* c^\dagger_{-k-Q}c^\dagger_{k-Q}$ of the  {\it same} type as the ones characterizing the ground state in the $S_p$ sector.
Finally for the $S_{p}$ sector
\begin{eqnarray}
|ex_3(k;Q)\rangle&=&    \prod_{{\tiny \begin{array}{c}   k'>0 \\ k' \in S_{p}\\ k' \neq k\end{array}}} \!\!\! (u_{k'} +v_{k'}^* \, c^{\dagger}_{-k'-Q}c^{\dagger}_{k'-Q}  ) \,\times \nonumber \\
& & \times \left( \prod_{ k' \in S_{2} }\,  c^{\dagger}_{k'-Q}\right) c^{\dagger}_{k-Q}  \,\,  
 \ket{0}     \label{ex-3}
\end{eqnarray}
where, as compared to the ground state (\ref{G-gen}), a Cooper pair in the $S_p$ sector  has been replaced by one single fermion.
\section{Lattice model}
\label{AppD}
\begin{figure}[h]
		\includegraphics[scale = 0.5]{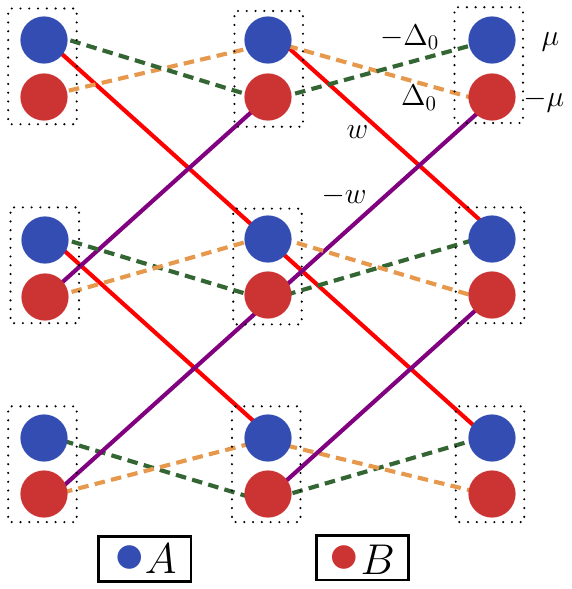}
		\caption{\label{Fig9}   Sketch of the various hopping terms of the 2D Hamiltonian (\ref{H2D-real-space}) defined in the  2D lattice. Here $A$ and $B$  represent either two orbitals  in the same physical lattice site, or two sites in the same unit cell. }
\end{figure}

In this Appendix, we provide additional details about the 2D lattice model described by the Hamiltonian \eqref{H2D}. The Hamiltonian $H(\bm{k})$ contained therein   can be written in the form 
\begin{widetext}
\begin{equation}
    H(\mathbf{k})  =\frac{1}{2}\left(\begin{array}{cc}
we^{i\left(Q-k\right)}+we^{-i\left(Q-k\right)}+\mu & -\Delta_{0}e^{ik}+\Delta_{0}e^{-ik}\\
\Delta_{0}e^{ik}-\Delta_{0}e^{-ik} & -we^{i\left(Q-k\right)}-we^{-i\left(Q-k\right)}-\mu
\end{array}\right)\quad.
\end{equation}
\end{widetext}
After Fourier transforming to real space, one obtains
\begin{eqnarray}
    \mathcal{H}_{2D} &=&\sum_{m,n}\Bigg\{\frac{\Delta_{0}}{2}\Big[f_{A\left(n+1,m\right)}^{\dagger}f_{B\left(n,m\right)}+f_{B\left(n,m\right)}^{\dagger}f_{A\left(n+1,m\right)}
    \nonumber\\
    &&-f_{A\left(n,m\right)}^{\dagger} f_{B\left(n+1,m\right)}-f_{A\left(n+1,m\right)}^{\dagger} f_{B\left(n,m\right)}\Big] \nonumber\\
	&& +\frac{w}{2}\Big[f_{A\left(n,m+1\right)}^{\dagger}f_{A\left(n+1,m\right)}+f_{A\left(n+1,m\right)}^{\dagger}f_{A\left(n,m+1\right)} \nonumber\\
    &&-f_{B\left(n+1,m+1\right)}^{\dagger}f_{B\left(n,m\right)}-f_{B\left(n,m\right)}^{\dagger}f_{B\left(n+1,m+1\right)}\Big] \nonumber\\
	&& +\frac{\mu}{2}\left[f_{A\left(n,m\right)}^{\dagger}f_{A\left(n,m\right)}-f_{B\left(n,m\right)}^{\dagger}f_{B\left(n,m\right)}\right]\Bigg\} \; \label{H2D-real-space}.
\end{eqnarray}
and the various terms are sketched in Fig.\ref{Fig9}. The $A$ and $B$ spots represent two different orbitals in the same site or two different sites in the same unit  cell. 
%

Although we are presently unaware of a material  described by the above Hamiltonian, the recent proposals of realizations of 2D WSMs with ultracold atoms\cite{guo2019}, suggest that its realization   in the near future    with synthetic materials, e.g., cold-atomic or photonic lattices  could be  possible.

\section{Fermi arcs}
\label{AppE}
In this Appendix, we consider the 2D Weyl Hamiltonian in Sec. \ref{sec4} with a boundary in the direction $x$, say, at $x=0$, while remaining infinite in the direction $y$. We look for the surface states, and aim in particular at determining their support in the surface Brillouin zone. To this end, we will focus on the low energy Hamiltonian expanding around the Weyl nodes.
In the Type-I phase, in order to describe the ``arc" joining the projection of the Weyl nodes $\bm{k}_W^{(0,\pm)}$, one obtains
\begin{equation}\label{H0}
    H_{0}(k;Q)	\approx	w\sin Q\, k\sigma_{0}-\Delta_{0}k \,\sigma_2+\left(w\cos Q-\frac{\mu}{2}\right) \sigma_3 \;,
\end{equation}
which contains the first term in the momentum $k$ expansion of \eqref{H2D} around $k=0$.  The same expansion is used in the Type-II phase, but the arc joins now the electron/hole pockets enclosing the nodes.
The most general boundary condition ensuring the self-adjointness of \eqref{H0} can be derived by taking two arbitrary spinors $\psi$ and $\chi$ and imposing that $\left\langle H\psi,\chi\right\rangle = \left\langle \psi,H\chi\right\rangle$
in the chosen geometry \cite{Witten2016}. In the absence of the phase modulation $Q$, this would imply the condition normally derived for Type-I Weyl semimetals \cite{Buccheri2022}. Here instead, we obtain the constraint
\begin{equation}\label{selfadjointness}
    w\sin Q\, \psi^{\dagger}\left(0\right)\chi\left(0\right)=\Delta_{0}\psi^{\dagger}\left(0\right)\sigma^{y}\chi\left(0\right) \;,
\end{equation}
provided $\psi$ and $\chi$ are well-behaved for $x\to\infty$. We note that, taking $\psi=\chi$, the above equation implies that the density current  $J_{x}= {\hbar}^{-1}\partial_{k}H$ across the boundary  vanishes. It also implies that the state on the surface must take the form
\begin{equation}\label{ssform}
     \chi_Q\left(x=0\right)=\frac{1}{\sqrt{2}}\left(\begin{array}{c}
e^{-i\zeta/2}\\
e^{i\zeta/2}
\end{array}\right)\quad,
\end{equation}
where the phase $\zeta$ must be a function of $Q$ and satisfy
\begin{equation}\label{wzeta}
    \frac{w}{\Delta_{0}}\sin Q=\sin\zeta \;.
\end{equation}
The latter    equation directly implies the  expression of $\zeta$ given in Sec. \ref{subsec:Lifshitz} and, for $Q=0$, reduces to the eigenstate of $\sigma^x$ with positive eigenvalue. This is the familiar form describing a straight ``arc" in Type-I Weyl semimetals or the Majorana end mode in the Kitaev model.
Taking an Ansatz wavefunction in the form \eqref{psi0} with an unspecified  $\lambda_Q$ in the exponent, it can be directly verified applying the Hamiltonian \eqref{H0} that this represents a zero-energy eigenstate only if the  penetration length is a function of the momentum $Q$ and the condition
\begin{equation}\label{cond1}
    w\cos Q-\frac{\mu}{2}-\frac{\Delta_{0}}{\lambda_Q}\cos\zeta=0
\end{equation}
is satisfied. Finally one solves for the penetration length $\lambda_Q$ and  the relation  \mbox{$\cos\zeta=\pm\sqrt{\Delta_0^2-w^2\sin^2Q}$}  selects the branch of the solution for $\zeta$ from the condition that $\lambda_Q>0$ along the arc. To this end, we recall that, in the Weyl-I phase, the arc connects the projection of the nodes $\bm{k}_W^{(0,\pm)}$ through the origin for $\mu>0$, i.e., $\cos Q>0$, while for $\mu<0$ the arc winds instead across the BZ and $\cos Q<0$. 
The same procedure can be readily applied to the Type-II Weyl phase, with the important difference that there appear bulk states around the Weyl nodes. Therefore, the arc support is shorter and spans the interval $\left[-Q^*,Q^*\right]$ for $\mu>0$, while it joins $\pi-Q^*$ and $-\pi+Q^*$ including the point $Q=\pi$ for $\mu<0$.

\begin{figure}[h]
    \centering
    \includegraphics[width=\columnwidth]{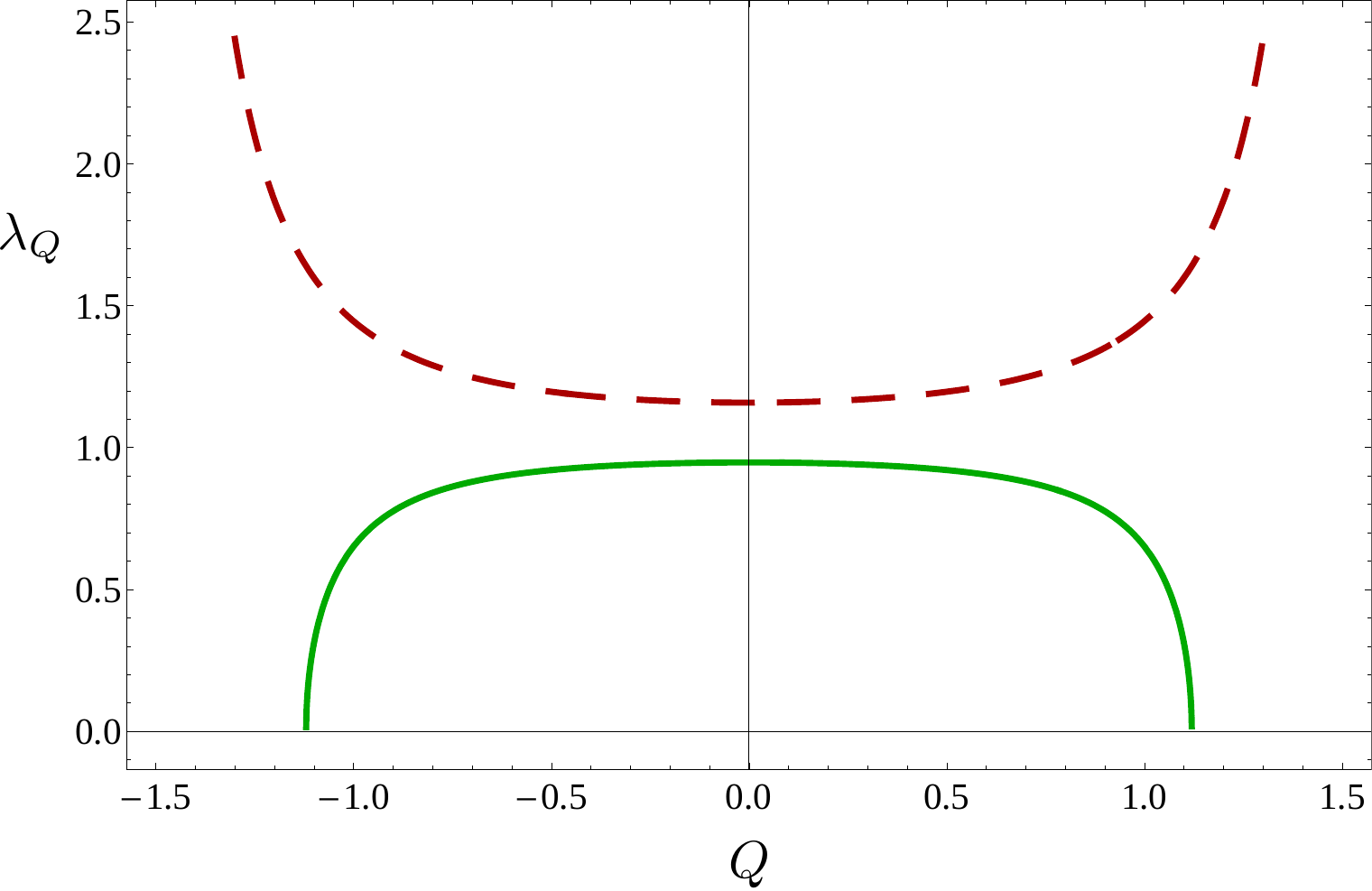}
    \caption{Plot of the penetration length $\lambda_Q$ in \eqref{lambda} of the arc including the origin in the Weyl-I and in the Weyl-II phase. The curves are obtained with parameters $\mu/w=0.1$ and $\Delta_0/w=1.1$ (dashed red line) and $\Delta_0/w=0.9$ (solid green line). \label{Fig10}}
\end{figure}

Similarly, one can describe the ``arc"   connecting  the projection of the Weyl nodes $\bm{k}_W^{(\pi,\pm)}$ by considering the expansion of \eqref{H2D} around $k=\pi$. It is readily verified that this Hamiltonian can be  formally obtained from $H_0$ by exchanging $w\to-w$ and $\Delta_0\to-\Delta_0$. By repeating the steps above, the surface state must once again reduce to \eqref{ssform} on the boundary and the form \eqref{psi0} describes a zero-energy state under imposing 
\begin{equation}\label{cond2}
    w\cos Q+\frac{\mu}{2}-\frac{ \Delta_{0}}{\lambda_Q}\cos\zeta=0\;.
\end{equation}
Solving \eqref{cond1} and \eqref{cond2} for  $\lambda_Q$, the branch of the phase $\zeta$ is again selected by the condition $\lambda_Q>0$.
To this end, it is important to notice that the Fermi arc support includes $Q=\pi$ if $\mu>0$, while it includes  the origin $Q=0$ for $\mu<0$.
Joining the above considerations, one finds the expression \eqref{lambda} in the main text, also represented in Fig. \ref{Fig10} for two sample values of parameters in the two Weyl phases.\\

\section{Analytical expression for the current at $\mu=0$}
\label{AppF}
Here we show that, starting from the general expression (\ref{current}) for the current, for the special case $\mu=0$ we can derive an analytical expression of $I(Q)$ that holds for arbitrary values of $\Delta_0$.
We first observe that, for $\mu=0$, the function (\ref{Phi-def}) determining gapless and gapped region reduces to $\Phi(k;Q)=\cos^2{Q}-(1-\Delta_0^2/w^2) \sin^2{k}$. Then, taking into account the sign of $h_0$ [see Eq.(\ref{h0-def})], the conditions   (\ref{eta-sectors-rel}) enable one to  determine  the values of the spectral asymmetry over the Brillouin zone
\begin{equation}
 \left\{ \begin{array}{l} 
   k^*_Q < k < \pi - k^*_Q  \,\,\rightarrow \,\,  \left\{ \begin{array}{lcl} \eta=+1 & \mbox{if} & Q \in ]0;\pi] \\
 \eta=-1 & \mbox{if} & Q \in [-\pi;0[
 \end{array} \right.\\   \\
 -\pi + k^*_Q < k < - k^*_Q  \,\,\rightarrow \,\,   \left\{ \begin{array}{lcl} \eta=-1 & \mbox{if} & Q \in ]0;\pi] \\
 \eta=+1 & \mbox{if} & Q \in [-\pi;0[
 \end{array} \right.\\    \\
  |k| <  k^*_Q \,\, \mbox{ or } \,\, \pi-k^*_Q<|k| < \pi  \,\,\,\rightarrow    \,\,\,  \eta=0\,\,   
\end{array}
\right. \label{eta-ranges-mu=0}
\end{equation}
where
\begin{eqnarray}
k^*_Q=\mbox{arcsin}\left( \frac{|\cos{Q}|}{\sqrt{1-(\Delta_0/w)^2}} \right)\,\,. 
\label{kstar-KTV}
\end{eqnarray}
The  expression (\ref{current}) of the current consists of two contributions, one  proportional to $\eta$, which represents the current $I_U(Q)$ carried by unpaired fermions and holes, and a second termn proportional to $1-|\eta|$, that represents the supercurrent  $I_S(Q)$ due to Cooper pairs. Let us start by determining   the former contribution. Indentifying through Eq.(\ref{eta-ranges-mu=0}) the $k$-regions where $\eta=\pm 1$, one finds 
\begin{eqnarray}
\lefteqn{I_U(Q;\mu = 0) =  - \frac{w\text{e}}{2\pi\hbar} \intop^{\pi}_{-\pi} dk \sin\left(k -Q\right)\eta\left(k , Q\right) =} & & \\
&=&   -4\frac{w\text{e}}{2\pi\hbar} \theta\left(\left|\sin\left(Q\right)\right| - \frac{\Delta_{0}}{w}\right) \cos\left(k^*_Q\right)\cos\left(Q\right)\text{sgn}\left( \sin {Q}\right)\nonumber
\end{eqnarray}
Recalling   Eq.(\ref{kstar-KTV}), one  obtains
\begin{eqnarray}
\lefteqn{I_U(Q;\mu = 0)=- \frac{4 w\text{e}}{2\pi\hbar} \theta\left(w\left|\sin\left(Q\right)\right| -  \Delta_{0} \right) \times } & & \nonumber \\ & & \times \sqrt{ \frac{w^2 \sin^2{Q}- \Delta_0^2}{ w^2-\Delta_0^2} } \cos\left(Q\right)\text{sgn}\left( \sin{Q}\right) \label{jU-mu=0}
\end{eqnarray}
Let us now consider the current associated to Cooper pairs ($\eta=0$).  From the expression (\ref{current_continuum-v0}), the current associated to the Cooper pairs is
\begin{eqnarray}
I_S(Q;\mu = 0)  &=& - \frac{w\text{e}}{2\pi \hbar}\intop^{\pi}_{-\pi} dk  \frac{ \sin\left(k -Q\right)\xi(k;Q)}{ \sqrt{ \xi^{2}(k;Q) + \left|\Delta_{k}\right|^{2}  }      } \times \nonumber \\
& & \hspace{1.5cm} \times \left(1 - \left|\eta\left(k,Q\right)\right|\right) \quad.
\end{eqnarray}
By introducing the quantity
\begin{equation}\label{delta-Q-def}
\delta_{Q} = \left(\frac{\Delta_{0}}{w\cos\left(Q\right)}\right)^{2} 
\end{equation}
and by exploiting the fact  that $|\eta(k,Q)|$ and $\xi(k,Q)$ are even functions of $k$, some straightforward algebra leads one to   write 
\begin{eqnarray}
\lefteqn{I_S(Q;\mu = 0)=  \frac{w\text{e}}{2\pi \hbar}\, 2\sin{Q}  \,\,\text{sgn}(\cos{Q}) \times} & & \nonumber \\
& & \times \intop^{\pi}_{0} dk  \frac{ \cos^2 k  }{ \sqrt{  \cos^2{k} + \delta_Q \sin^2{k}   } } \left(1 - \left|\eta\left(k,Q\right)\right|\right) \label{jS-mu=0-pre}
\end{eqnarray}
One can now identify  from  Eq.(\ref{eta-ranges-mu=0})  the regions with $\eta=0$ and write 
\begin{widetext}
\begin{eqnarray}
I_S(Q;\mu = 0) &=&   \frac{w\text{e}}{2\pi \hbar}\, 2\sin Q  \,\,\text{sgn}(\cos Q)  \left\{\theta\left(w |\sin  {Q} | -  \Delta_{0} \right)   \left(  \intop_{0}^{k^*_Q} dk  \frac{ \cos^2 k  }{ \sqrt{  \cos^2{k} + \delta_Q \sin^2{k}   } }  +\intop^{\pi}_{\pi-k^*_Q} dk  \frac{ \cos^2 k  }{ \sqrt{  \cos^2{k} + \delta_Q \sin^2{k}   } }  \right)-\right. \nonumber \\
& &  \hspace{4cm}\left.-   \, \theta\left( \Delta_{0}  -w|\sin {Q}|   \right)\intop^{\pi}_{0} dk  \frac{ \cos^2 k  }{ \sqrt{  \cos^2{k} + \delta_Q \sin^2{k}   } }  \right\}
\end{eqnarray}
Let us now focus on the two integrals appearing in the first term. We notice that, by changing the integration variable $k=\pi-k^\prime$ in the second integral, the latter turns out to be equal to the first one. Similarly, the integral appearing in the second term can be written as twice the integral from $0$ to $\pi/2$. This enables one to rewrite 
\begin{eqnarray}
I_S(Q;\mu = 0)  &=&     \frac{w\text{e}}{2\pi \hbar}\, \frac{4\sin{Q}  \,\,\text{sgn}(\cos{Q}) }{1-\delta_Q}    \left\{\theta\left(\left|\sin\left(Q \right)\right| - \frac{\Delta_{0}}{w}\right) \left(  E(k^*_Q;1-\delta_Q)-\delta_\phi\,\, F(k^*_Q;1-\delta_Q) \right) \right. \nonumber\\
& & \left. \hspace{3.5cm} +\theta\left( \frac{\Delta_{0}}{w} -\left|\sin\left(Q \right)\right|  \right) \sqrt{\delta_Q} \left(  E(1-\frac{1}{\delta_Q})- \, K(1-\frac{1}{\delta_Q}) \right) \right\} \label{jS-mu=0}
\end{eqnarray}
where   $E$, $F$ and $K$ are incomplete elliptic functions, while $E(x) \equiv E(\frac{\pi}{2};x)$ and $ K(x) \equiv F(\frac{\pi}{2};x)$. Finally, exploiting the properties of the Elliptic integrals
$
\sqrt{x} E\left(1-1/x\right)  = E(1-x)$ and $ 
\sqrt{x} K\left(1-1/x\right) = x\, K(1-x)$ for $x>0$, and 
combining together the contributions (\ref{jU-mu=0}) and (\ref{jS-mu=0}), one finally obtains the current at $\mu = 0$ 
\begin{equation}
\begin{array}{lcl}
I\left(Q;\mu=0\right)  &=&-
\displaystyle \frac{4w\text{e}}{2\pi\hbar} \left\{ \theta\left(w\left|\sin\left(Q\right)\right| -  \Delta_{0} \right) \left[ \sqrt{ \frac{w^2 \sin^2{Q}- \Delta_0^2}{ w^2-\Delta_0^2} } \cos\left(Q\right)\text{sgn}\left( \sin Q \right)  -  \right. \right.\\
& & \hspace{2cm} \displaystyle \left.     -\frac{ \sin Q  \,\,\text{sgn}(\cos Q) }{1-\delta_Q} \left(  E(k^*_Q;1-\delta_Q )-\delta_Q\,\, F(k^*_Q;1-\delta_Q) \right) \right]- \\ & & \\
& & \displaystyle \left. \hspace{1cm} -\theta\left(  \Delta_{0}  -w\left|\sin Q\right|  \right) \frac{ \sin Q  \,\,\text{sgn}(\cos Q) }{1-\delta_Q} \left(  E(1-\delta_Q)-\delta_Q\,\, K(1-\delta_Q) \right) \right\}
\end{array}
\label{jTOT-mu=0}
\end{equation}
where $\delta_Q$ is Eq.(\ref{delta-Q-def}), and $k^*_Q$ is Eq.(\ref{kstar-KTV}). The current near $Q=0$ and for $\Delta_0 \ll w$ gives
\begin{equation}
I(Q;\mu=0) \sim \frac{w {\rm e} \, Q}{2\pi \hbar}\left( 4+\left(3 -4\ln{2}+2 \ln\left( \frac{\Delta_0}{w}\right) \right)\frac{\Delta^2_0}{w^2}\right) 
\end{equation}
and to leading order the slope $dI/dQ$   depends only on the bandwidth parameter $w$.

\end{widetext} 

	\bibliography{Biblio}

\end{document}